\documentclass[prd,letterpaper,twocolumn,preprintnumbers,nofootinbib]{revtex4}
\usepackage[letterpaper, hdivide={1.91cm,,1.165cm}, vdivide={1.83cm,,2.5cm}]{geometry}

\usepackage{slashed}
\usepackage{amsmath,amssymb}
\usepackage{graphicx}
\usepackage{units}
\usepackage{bbold}
\usepackage{xcolor}
\usepackage{dsfont}
\usepackage[hyperfootnotes=false,colorlinks,citecolor=blue]{hyperref}
\usepackage{float}
\usepackage{subcaption}
\usepackage{array}
\usepackage{pifont}

\usepackage[normalem]{ulem}

\def \vec#1{{\boldsymbol{#1}}}

\newcommand{\hc}{\ensuremath{\text{h.c.}}}

\newcommand{\ii}{{\rm i}}

\def \check{\ding{52}}
\newcommand{\cross}{\ding{56}}

\begin{document}

\title{Lepton flavor violation at mass dimension eight}

\author{Julian Heeck}
\email[Email: ]{heeck@virginia.edu}
\thanks{ORCID: \href{https://orcid.org/0000-0003-2653-5962}{0000-0003-2653-5962}.}
\affiliation{Department of Physics, University of Virginia,
Charlottesville, Virginia 22904-4714, USA}

\author{Mikheil Sokhashvili}
\email[Email: ]{ms2guc@virginia.edu}
\thanks{ORCID: \href{https://orcid.org/0000-0003-0844-7563}{0000-0003-0844-7563}.}
\affiliation{Department of Physics, University of Virginia,
Charlottesville, Virginia 22904-4714, USA}

\hypersetup{
pdftitle={Lepton flavor violation at mass dimension eight},   
pdfauthor={Julian Heeck, Mikheil Sokhashvili}
}

\begin{abstract}

Operators that violate lepton flavor first arise at mass dimension $d=5$ in the Standard Model Effective Field Theory, while \emph{charged} lepton flavor violation first appears at $d=6$, inducing the typical exotic particle decays currently probed by experiments. Operators at $d=8$ carry the same quantum numbers as those at $d=6$ and are thus seemingly doomed to be subleading corrections. After (re)deriving the complete set of tree-level UV completions for lepton-flavor-violating $d\leq 8$ operators via heavy scalars and fermions, we show that some UV completions exclusively generate $d=8$ flavor violation at tree level, with $d=6$ contributions suppressed by loop factors and, in some cases, unrelated couplings. Such accidentally protected $d=8$ scenarios enlarge the set of testable lepton-flavor-violating operators.

\end{abstract}

\maketitle

\section{Introduction}
\label{sec:intro}
\vspace{-1ex}

The immensely successful Standard Model of particle physics (SM) predicts the conservation of baryon, electron, muon, and tauon number, connected by Noether's theorem to four $U(1)$ symmetries: $U(1)_B\times U(1)_{L_e}\times U(1)_{L_\mu}\times U(1)_{L_\tau}$. These are often dubbed \emph{accidental} symmetries, since they are not imposed on the SM Lagrangian but rather emerge as a result of the SM field content (Tab.~\ref{tab:fields}) and the restriction to renormalizable terms.
The fate of these global symmetries is unknown: they might be violated through quantum gravity or due to new particles, but (some linear combinations) could be strictly conserved after all~\cite{Heeck:2014zfa}. It is ultimately an experimental question, one that has motivated many dedicated experiments to search for baryon number~\cite{FileviezPerez:2022ypk} or lepton flavor violation (LFV)~\cite{Davidson:2022jai}, aided greatly by the fact that such searches are essentially free of background.

LFV is particularly well motivated since the observation of neutrino oscillations proves that lepton flavor is not conserved in nature, and that the SM is incomplete and in need of additional particles. 
Assuming all new particles beyond the SM are heavy, we can organize LFV using effective operators in the Standard Model Effective Field Theory (SMEFT). The lowest-dimensional non-renormalizable operator that can be written down using the SM fields from Tab.~\ref{tab:fields} already violates lepton flavor: this Weinberg operator $HHLL$~\cite{Weinberg:1979sa} also breaks \emph{total} lepton number and induces Majorana neutrino masses, providing a simple explanation for all neutrino oscillation data. \emph{Charged} LFV is induced as well, but heavily suppressed by the small neutrino masses, rendering processes such as $\mu\to e\gamma$ completely unobservable. 

At mass dimension $d=6$, one finds LFV SMEFT operators that do not violate total lepton number and are hence generically decoupled from neutrino masses; many of these $d=6$ operators could induce observable LFV rates and provide the dominant inspiration for LFV signatures~\cite{Davidson:2022jai}. LFV operators with $\Delta L_\mu = -\Delta L_e = 1$ are subject to the strongest experimental constraints, while $\Delta L_\tau=2$ operators are weakly constrained, if at all, as recently emphasized in Ref.~\cite{Heeck:2024uiz}.
At $d=7$, LFV once again comes with total lepton number or even baryon number violation, but could still induce testable signatures unsuppressed by neutrino masses~\cite{Heeck:2025jfs}.

\begin{table}[tb]
    \renewcommand{\arraystretch}{1.3} 
\begin{tabular}{c c c c} 
 \hline
 Field & Chirality & Generations & $SU(3)_C\times SU(2)_L\times U(1)_Y$   \\ [0.5ex] 
 \hline\hline
 $Q$ & left & 3 & $\left(\vec{3},\vec{2},\tfrac16\right)$\\
 $u$ & right & 3 & $\left(\vec{3},\vec{1},\tfrac23\right)$\\
 $d$ & right & 3 & $\left(\vec{3},\vec{1},-\tfrac13\right)$\\
 $L$ & left & 3 & $\left(\vec{1},\vec{2},-\tfrac12\right)$\\
 $e$ & right & 3 & $\left(\vec{1},\vec{1},-1\right)$\\
 $H$ & scalar & 1 & $\left(\vec{1},\vec{2},\tfrac12\right)$\\
 \hline
\end{tabular}
\caption{SM fields and quantum numbers; hypercharge is related to electric charge via $Q = Y + T_3$ with diagonal $SU(2)_L$ generator $T_3$.}
\vspace{-2ex}
\label{tab:fields}
\end{table}

Pure \emph{charged} LFV again arises at $d=8$, with only a few phenomenological studies in the literature so far. Given the extraordinary experimental sensitivity to $\Delta L_\mu = -\Delta L_e = 1$ processes, dimension-eight operators could absolutely induce testable LFV rates~\cite{Ardu:2021koz}; however, since the $d=8$ operators~\cite{Li:2020gnx,Murphy:2020rsh} have the same quantum numbers as the $d=6$ ones, one cannot impose a simple flavor symmetry as in Refs.~\cite{Heeck:2016xwg,Heeck:2024uiz,Heeck:2025jfs} to make them the dominant LFV operators. At first sight, this makes the $d=8$ LFV operators rather uninteresting, essentially subleading corrections to the not-yet-observed $d=6$ signatures. 
However, there are some $d=8$ operators that generate interesting novel signatures that might give them an edge over $d=6$:
Refs.~\cite{PhysRev.126.375,Bowman:1978kz,Davidson:2020ord,Bryman:2021ilc,Fortuna:2022sxt,Fortuna:2023paj} considered dimension-eight LFV operators involving two electromagnetic field-strength tensors, e.g.~$\bar{L} H e\, F_{\mu\nu}F^{\mu\nu}$, and 
Refs.~\cite{Petrov:2013vka,Bhattacharya:2018ryy,Cai:2018cog} similar operators involving two gluonic field-strength tensors, all arising from loop-level UV completions.
Beyond those studies, Ref.~\cite{Ardu:2021koz} considered the impact of other $d=8$ operators on charged LFV in a particular power-counting scheme, notably highlighting $d=8$ SMEFT operators that lead to novel chirality structures after electroweak symmetry breaking, which in principle allows experimental distinction of $d=6$ and $d=8$ LFV operators.

Here, we approach dimension-eight LFV from a different angle. Since $d=6$ and $d=8$ carry the same quantum numbers, SMEFT provides no guidance on how to make $d=8$ dominant. We therefore go beyond SMEFT and study \emph{UV completions} of $d=8$ operators to systematically search for those that do not generate any LFV at $d<8$. While not protected by any symmetry, this provides an \emph{accidental} protection of $d=8$ operators, i.e.~a mechanism to make some $d=8$ operators the starting point for LFV. $d=6$ operators will inevitably be generated via loops,\footnote{See also Ref.~\cite{Cepedello:2022pyx} for a related exploration of models that suppress $d=6$ four-fermion operators to loop level.} but will not necessarily dominate the phenomenology, depending on the model parameters. Even if the loops dominate this is interesting, as our UV completions furnish genuine/finite loop realizations of $d=6$ operators, with characteristic Wilson-coefficient ratios.

Our approach requires tree-level UV completions for $d\leq 8$ LFV SMEFT operators, which have been collected in other articles already. We nevertheless re-derived all UV completions via heavy fermions and scalars in weakly-coupled renormalizable theories using the code from Refs.~\cite{Heeck:2026dmh,Heeck:2026lhd}. The UV completions for $d=5$ are collected in Sec.~\ref{sec:UV5}, those for $d=6$ in Sec.~\ref{sec:UV6}, and $d=7$ in Sec.~\ref{sec:UV7}. To keep the article length manageable, we do not list \emph{all} $d=8$ UV completions, but rather provide them as ancillary files to the arXiv version of this article~\cite{anc_files}. Instead, we focus on $d=8$ UV completions that by themselves do not induce LFV operators at $d<8$ and discuss their phenomenology in Sec.~\ref{sec:UV8}.
We conclude in Sec.~\ref{sec:conclusion}. Notice that we are concerned with LFV here, not just \emph{charged} LFV, so our analysis includes neutrino processes, unlike, for example, Ref.~\cite{Ardu:2021koz}.

\section{\texorpdfstring{$d = 5$}{d = 5}}
\label{sec:UV5}

\begin{figure}[tb]
    \begin{subfigure}[b]{0.2\textwidth}
        \includegraphics[width=\textwidth]{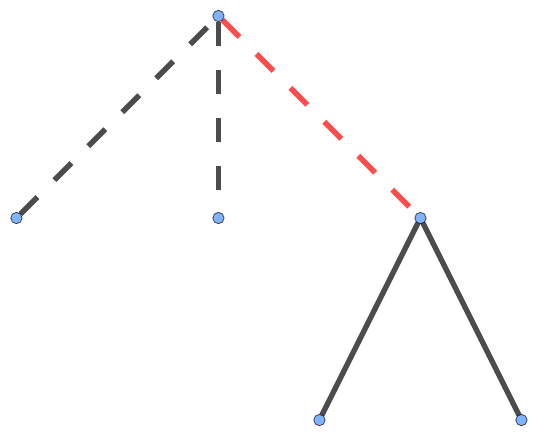}
        \caption{}
        \label{Topo_5_1}
    \end{subfigure}
    \hspace{0.05\textwidth}
    \begin{subfigure}[b]{0.2\textwidth}
        \includegraphics[width=\textwidth]{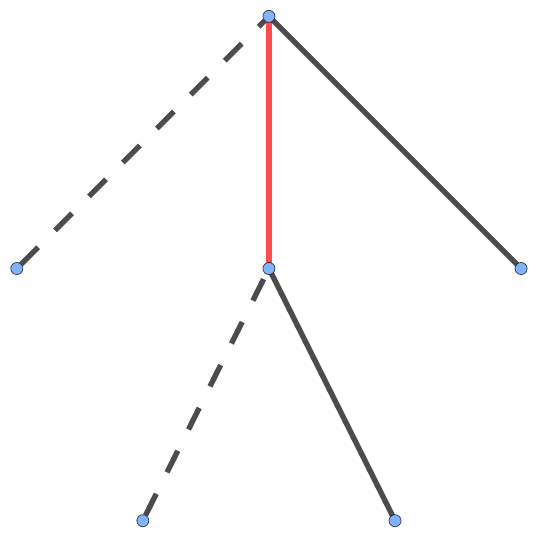}
        \caption{}
        \label{Topo_5_2}
    \end{subfigure}
    \caption{Feynman-diagram topologies for tree-level UV completions of $d=5$ LFV operators.}
    \label{fig:dim5_topos}
\end{figure}

At mass dimension $d = 5$, there exists a single SMEFT operator that violates lepton flavor, and also lepton number: the famous Weinberg operator~\cite{Weinberg:1979sa}:
\begin{align}
    \mathbb{O}_5 = \{HHLL\} \,,
\end{align}
where we used a simplified notation that suppresses all Lorentz and gauge group contractions and only keeps the involved particles, sometimes referred to as \emph{field strings} or \emph{operator types}. 
The corresponding (tree-level) UV completions are particularly well studied and are commonly referred to as seesaw mechanisms: type-I~\cite{Minkowski:1977sc,Yanagida:1979as,Gell-Mann:1979vob,Mohapatra:1979ia}, type-II~\cite{Magg:1980ut,Schechter:1980gr,Wetterich:1981bx,Lazarides:1980nt,Mohapatra:1980yp,Cheng:1980qt}, and type-III~\cite{Foot:1988aq}. They arise from the two Feynman-diagram topologies shown in Fig.~\ref{fig:dim5_topos}. Here and below, dashed lines denote scalars and solid lines fermions, with red lines representing heavy mediator fields and black lines external SM fields.  

The UV completions or mediator particles are collected in Tab.~\ref{tab:d_5}. Each row corresponds to a distinct UV completion scenario, specifying the set of mediating particles that generate the given operator at tree-level. Within each row, mediators are denoted by ``$S$'' or ``$F$,'' referring to scalar or fermion fields, respectively, accompanied by their SM gauge-group representation in subscript.
Notice that we only have \emph{one} Higgs doublet in the SM, and therefore a fourth UV completion via $S_{(\mathbf{1}, \mathbf{1}, 1)}$ and topology~\ref{Topo_5_1} vanishes.
Here and in the following, all internal particles are assumed to be heavy, and fermions are hence vector-like/Dirac or potentially Majorana if their gauge representation is real.

\begin{table}[tb]
    \renewcommand{\arraystretch}{1.4} 
    \setlength{\tabcolsep}{10pt} 
    \begin{tabular}{|c|c|c|}
        \hline
        Operator & UV Completion & Topo. \\
        \hline \hline
        $HHLL$
        & $S_{(\mathbf{1}, \mathbf{3}, 1)}$ & Fig.~\ref{Topo_5_1}\\
        & $F_{(\mathbf{1}, \mathbf{3}, 0)}$ & Fig.~\ref{Topo_5_2}\\
        & $F_{(\mathbf{1}, \mathbf{1}, 0)}$ & Fig.~\ref{Topo_5_2}\\

        \hline
    \end{tabular}
    \caption{Tree-level UV completions for $d = 5$ LFV operators, with $S$ vs.~$F$ indicating scalar vs.~fermion and the last column indicating the appropriate topology.}
    \label{tab:d_5}
\end{table}

After electroweak symmetry breaking, the Weinberg operator generates Majorana neutrino masses and lepton-flavor-violating neutrino oscillations $\nu_\alpha\to \nu_\beta$. These have been observed and could be taken as evidence for the Weinberg operator if it were not for the fact that the qualitatively different scenario of Dirac neutrinos is equally valid. In any case, $HHLL$ and its UV completions induce LFV, which has been studied extensively in the literature. \emph{Charged} LFV is generically suppressed since $HHLL\to \langle H\rangle^2 \nu\nu$ only contains neutrinos, requiring $W$-boson interactions to end up with charged leptons. 
Aside from maybe neutrinoless double-beta decay~\cite{Rodejohann:2011mu}, we hence do not expect any LFV signatures at $d=5$ aside from neutrino oscillations.

\section{\texorpdfstring{$d = 6$}{d = 6}}
\label{sec:UV6}

At mass dimension $d=6$, there exist numerous SMEFT operators that violate lepton flavor, and only some also violate total lepton and baryon number:
\begin{align}
\begin{split}
    \mathbb{O}_6 =
    \{
    &\bar{Q}\bar{L} eu, \bar{Q}\bar{L}LQ, \bar{Q}\bar{ e} eQ , \bar{Q}\bar{ e}Ld , \bar{d}\bar{L}Ld, \bar{d}\bar{ e} ed, \\
    & \bar{u}\bar{L}Lu, \bar{u}\bar{ e} eu, \bar{L}\bar{L}LL, \bar{L}\bar{ e} eL, \bar{ e}\bar{ e} e e, euud, \\
    & euQQ, LudQ, LQQQ, \bar{H}\bar{L} eHH, \\
    &  \mathcal{D} \bar{H} H \bar{L} L, \mathcal{D} \bar{H} H \bar{e} e, \bar{L}  e H W, \bar{L}  e H B
    \} \,.
\end{split}
\end{align}
Here, $B$ and $W$ abstractly denote the $U(1)_Y$ and $SU(2)_L$ field strength tensors, and $\mathcal{D}$ the covariant derivative.
A proper operator basis with well-defined index contractions has been derived in Ref.~\cite{Grzadkowski:2010es} but is not needed for our purposes.

\begin{figure}[tb]
    \centering
    \begin{subfigure}[b]{0.4\columnwidth}
        \includegraphics[width=\textwidth]{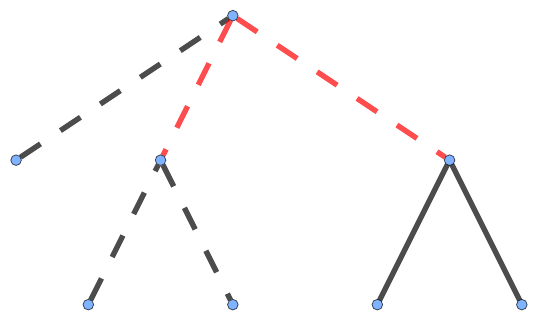}
        \caption{}
        \label{Topo_6_1}
    \end{subfigure}
    \hfill
    \begin{subfigure}[b]{0.4\columnwidth}
        \includegraphics[width=\textwidth]{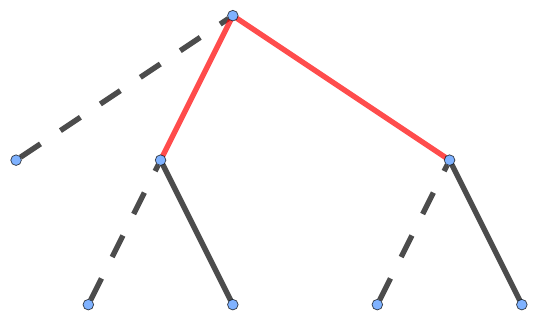}
        \caption{}
        \label{Topo_6_2}
    \end{subfigure}

    \vspace{0.5em} 

    \begin{subfigure}[b]{0.4\columnwidth}
        \includegraphics[width=\textwidth]{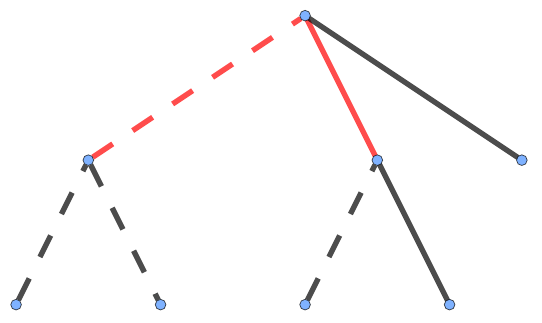}
        \caption{}
        \label{Topo_6_3}
    \end{subfigure}
    \hfill
    \begin{subfigure}[b]{0.4\columnwidth}
        \includegraphics[width=\textwidth]{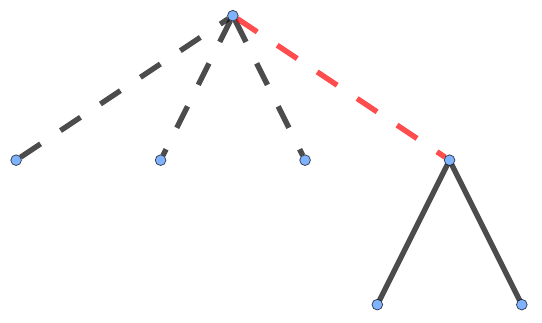}
        \caption{}
        \label{Topo_6_4}
    \end{subfigure}

    \vspace{0.5em} 

    \begin{subfigure}[b]{0.4\columnwidth}
        \includegraphics[width=\textwidth]{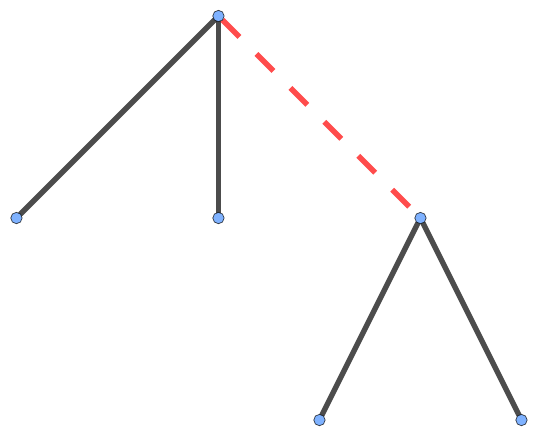}
        \caption{}
        \label{Topo_6_5}
    \end{subfigure}

    \caption{Topologies for tree-level UV completions of non-derivative $d=6$ LFV operators.}
    \label{fig:dim6_topos}
\end{figure}

Tree-level UV completions for these dimension-6 LFV operators are provided in Refs.~\cite{deBlas:2017xtg,Li:2022abx}. We also checked the results using the relevant topologies given in Fig.~\ref{fig:dim6_topos} and the code developed in Refs.~\cite{Heeck:2026dmh,Heeck:2026lhd} and find perfect agreement with the literature. 
The UV completions for non-derivative and derivative operators are collected in Tabs.~\ref{tab:d_6} and~\ref{tab:d_6_der}, respectively. Notice that operators involving field-strength tensors do not have tree-level UV completions and are hence not listed here.

The $d=6$ operator $e HH\bar{H}\bar{L}$ can seemingly also be UV-completed with two fermions $F_{(\vec{1},\vec{1},0)}$ \& $F_{(\vec{1},\vec{2},1/2)}$ via Fig.~\ref{Topo_6_2}, but even though the Feynman diagram is valid, the corresponding amplitude simply vanishes to lowest order in the heavy-mediator expansion because it is antisymmetric under exchange of the two identical $H$. Expanding the mediator propagator to second order in momenta gives a non-vanishing amplitude, so $F_{(\vec{1},\vec{1},0)}$ \& $F_{(\vec{1},\vec{2},1/2)}$ actually form a UV completion for the $d=8$ operator $\mathcal{D}\mathcal{D}e HH\bar{H}\bar{L}$. 

\begin{table}[H]
    \centering
    \renewcommand{\arraystretch}{1.3} 
    \setlength{\tabcolsep}{10pt} 
    \begin{tabular}{|c|c c|c|}
        \hline
        Operator & \multicolumn{2}{c|}{UV Completion}   & Topo. \\
        \hline \hline
        $eHH\bar{H}\bar{L}$
        & $S_{(\mathbf{1}, \mathbf{2}, \tfrac{1}{2})}$ & $S_{(\mathbf{1}, \mathbf{3}, 1)}$ & Fig.~\ref{Topo_6_1}\\
        & $S_{(\mathbf{1}, \mathbf{2}, \tfrac{1}{2})}$ & $S_{(\mathbf{1}, \mathbf{3}, 0)}$ & Fig.~\ref{Topo_6_1}\\
        & $S_{(\mathbf{1}, \mathbf{1}, 0)}$ & $S_{(\mathbf{1}, \mathbf{2}, \tfrac{1}{2})}$ & Fig.~\ref{Topo_6_1}\\
        & $F_{(\mathbf{1}, \mathbf{2}, \tfrac{1}{2})}$ & $F_{(\mathbf{1}, \mathbf{3}, 1)}$ & Fig.~\ref{Topo_6_2}\\
        & $F_{(\mathbf{1}, \mathbf{1}, 1)}$ & $F_{(\mathbf{1}, \mathbf{2}, \tfrac{1}{2})}$ & Fig.~\ref{Topo_6_2}\\
        & $F_{(\mathbf{1}, \mathbf{2}, \tfrac{1}{2})}$ & $F_{(\mathbf{1}, \mathbf{3}, 0)}$ & Fig.~\ref{Topo_6_2}\\
        & $F_{(\mathbf{1}, \mathbf{2}, \tfrac{3}{2})}$ & $F_{(\mathbf{1}, \mathbf{3}, 1)}$ & Fig.~\ref{Topo_6_2}\\
        & $F_{(\mathbf{1}, \mathbf{1}, 1)}$ & $F_{(\mathbf{1}, \mathbf{2}, \tfrac{3}{2})}$ & Fig.~\ref{Topo_6_2}\\
        & $F_{(\mathbf{1}, \mathbf{2}, \tfrac{1}{2})}$ & $S_{(\mathbf{1}, \mathbf{3}, 0)}$ & Fig.~\ref{Topo_6_3}\\
        & $F_{(\mathbf{1}, \mathbf{2}, \tfrac{1}{2})}$ & $S_{(\mathbf{1}, \mathbf{1}, 0)}$ & Fig.~\ref{Topo_6_3}\\
        & $F_{(\mathbf{1}, \mathbf{3}, 1)}$ & $S_{(\mathbf{1}, \mathbf{3}, 0)}$ & Fig.~\ref{Topo_6_3}\\
        & $F_{(\mathbf{1}, \mathbf{1}, 1)}$ & $S_{(\mathbf{1}, \mathbf{1}, 0)}$ & Fig.~\ref{Topo_6_3}\\
        & $F_{(\mathbf{1}, \mathbf{2}, \tfrac{3}{2})}$ & $S_{(\mathbf{1}, \mathbf{3}, 1)}$ & Fig.~\ref{Topo_6_3}\\
        & $F_{(\mathbf{1}, \mathbf{3}, 0)}$ & $S_{(\mathbf{1}, \mathbf{3}, 1)}$ & Fig.~\ref{Topo_6_3}\\
        & $S_{(\mathbf{1}, \mathbf{2}, \tfrac{1}{2})}$ & - & Fig.~\ref{Topo_6_4}\\
        \hline
        $e\bar{L}\bar{Q}u$
        & $S_{(\mathbf{3}, \mathbf{1}, -\tfrac{1}{3})}$ & - & Fig.~\ref{Topo_6_5}\\
        & $S_{(\mathbf{3}, \mathbf{2}, \tfrac{7}{6})}$ & - & Fig.~\ref{Topo_6_5}\\
        & $S_{(\mathbf{1}, \mathbf{2}, \tfrac{1}{2})}$ & - & Fig.~\ref{Topo_6_5}\\
        \hline
        $L\bar{L}Q\bar{Q}$
        & $S_{(\mathbf{3}, \mathbf{3}, -\tfrac{1}{3})}$ & - & Fig.~\ref{Topo_6_5}\\
        & $S_{(\mathbf{3}, \mathbf{1}, -\tfrac{1}{3})}$ & - & Fig.~\ref{Topo_6_5}\\
        \hline
        $\bar{e}eQ\bar{Q}$
        & $S_{(\mathbf{3}, \mathbf{2}, \tfrac{7}{6})}$ & - & Fig.~\ref{Topo_6_5}\\
        \hline
        $d\bar{e}L\bar{Q}$
        & $S_{(\mathbf{1}, \mathbf{2}, \tfrac{1}{2})}$ & - & Fig.~\ref{Topo_6_5}\\
        \hline
        $\bar{d}dL\bar{L}$
        & $S_{(\mathbf{3}, \mathbf{2}, \tfrac{1}{6})}$ & - & Fig.~\ref{Topo_6_5}\\
        \hline
        $\bar{d}d\bar{e}e$
        & $S_{(\mathbf{3}, \mathbf{1}, -\tfrac{4}{3})}$ & - & Fig.~\ref{Topo_6_5}\\
        \hline
        $L\bar{L}\bar{u}u$
        & $S_{(\mathbf{3}, \mathbf{2}, \tfrac{7}{6})}$ & - & Fig.~\ref{Topo_6_5}\\
        \hline
        $\bar{e}e\bar{u}u$
        & $S_{(\mathbf{3}, \mathbf{1}, -\tfrac{1}{3})}$ & - & Fig.~\ref{Topo_6_5}\\
        \hline
        $LL\bar{L}\bar{L}$
        & $S_{(\mathbf{1}, \mathbf{3}, 1)}$ & - & Fig.~\ref{Topo_6_5}\\
        & $S_{(\mathbf{1}, \mathbf{1}, 1)}$ & - & Fig.~\ref{Topo_6_5}\\
        \hline
        $\bar{e}eL\bar{L}$
        & $S_{(\mathbf{1}, \mathbf{2}, \tfrac{1}{2})}$ & - & Fig.~\ref{Topo_6_5}\\
        \hline
        $\bar{e}\bar{e}ee$
        & $S_{(\mathbf{1}, \mathbf{1}, 2)}$ & - & Fig.~\ref{Topo_6_5}\\
        \hline
        $deuu$
        & $S_{(\mathbf{3}, \mathbf{1}, -\tfrac{1}{3})}$ & - & Fig.~\ref{Topo_6_5}\\
        & $S_{(\mathbf{3}, \mathbf{1}, -\tfrac{4}{3})}$ & - & Fig.~\ref{Topo_6_5}\\
        \hline
        $eQQu$
        & $S_{(\mathbf{3}, \mathbf{1}, -\tfrac{1}{3})}$ & - & Fig.~\ref{Topo_6_5}\\
        \hline
        $dLQu$
        & $S_{(\mathbf{3}, \mathbf{1}, -\tfrac{1}{3})}$ & - & Fig.~\ref{Topo_6_5}\\
        \hline
        $LQQQ$
        & $S_{(\mathbf{3}, \mathbf{3}, -\tfrac{1}{3})}$ & - & Fig.~\ref{Topo_6_5}\\
        & $S_{(\mathbf{3}, \mathbf{1}, -\tfrac{1}{3})}$ & - & Fig.~\ref{Topo_6_5}\\
        \hline
    \end{tabular}
    \caption{Tree-level UV completions for non-derivative $d = 6$ LFV operators.}
    \label{tab:d_6}
\end{table}

Charged-LFV phenomenology is dominated by the $d=6$ operators, as these give LFV without connection to neutrino masses, thus potentially enhanced and measurable. Four of the operators further violate $\Delta B = 1 = \Delta L$ and thus induce proton decay~\cite{Weinberg:1979sa}; the other ones have been discussed extensively in the context of LFV, leading to decays such as $\ell \to \ell' \ell''\ell'''$, $\ell\to \ell' \gamma$, LFV decays involving mesons, Higgs,  $Z$ bosons, and more~\cite{Crivellin:2013hpa,Davidson:2022jai}. The most stringent limits come from muon and tauon decays, which require mapping of SMEFT operators onto the Low-Energy Effective Field Theory (LEFT), in which the heavy $Z$, $W$, Higgs, and top quark are integrated out~\cite{Jenkins:2017jig}; it is interesting to note that the $d=6$ SMEFT operators do not induce \emph{all} LFV-relevant $d=6$ LEFT operators at tree level, as some chirality combinations are missing. The remaining ones are induced by $d=8$ SMEFT operators instead~\cite{Ardu:2021koz,Hamoudou:2022tdn}. We shall come back to this in Sec.~\ref{sec:UV8}.

\begin{table}[tb]
    \centering
    \renewcommand{\arraystretch}{1.4} 
    \setlength{\tabcolsep}{10pt} 
    \begin{tabular}{|c|c|c|}
        \hline
        Operator & UV Completions & Topo. \\
        \hline \hline
        $\mathcal{D} \bar{H} H \bar{e} e$
        & $F_{(\mathbf{1}, \mathbf{2}, \tfrac{3}{2})}$ & Fig.~\ref{Topo_5_2}\\
        & $F_{(\mathbf{1}, \mathbf{2}, \tfrac{1}{2})}$ & Fig.~\ref{Topo_5_2}\\
        \hline
        $\mathcal{D} \bar{H} H \bar{L} L$
        & $F_{(\mathbf{1}, \mathbf{3}, 0)}$ & Fig.~\ref{Topo_5_2}\\
        & $F_{(\mathbf{1}, \mathbf{1}, 0)}$ & Fig.~\ref{Topo_5_2}\\
        & $F_{(\mathbf{1}, \mathbf{3}, 1)}$ & Fig.~\ref{Topo_5_2}\\
        & $F_{(\mathbf{1}, \mathbf{1}, 1)}$ & Fig.~\ref{Topo_5_2}\\
        \hline        
    \end{tabular}
    \caption{Tree-level UV completions for derivative $d = 6$ LFV operators.}
    \label{tab:d_6_der}
\end{table}

\section{\texorpdfstring{$d = 7$}{d = 7}}
\label{sec:UV7}

At $d=7$, all LFV operators additionally violate total lepton or baryon number:
\begin{align}
\begin{split}
    \mathbb{O}_7 =
    \{
    &\bar{H}\bar{L}udd, \bar{H}\bar{L}dQQ, \bar{H}\bar{e}ddQ, \bar{L}dddH, \bar{Q}LLuH,  \\
    & \bar{d}eLuH, \bar{d}LLQH, \bar{e}LLLH, \bar{H}LLHHH, \\
    & \mathcal{D} \bar{d} L L u, \mathcal{D} d d \bar{L} Q, \mathcal{D} d d d \bar{e}, \mathcal{D} \mathcal{D} H H L L, \\
    & H H L L B, H H L L W, \mathcal{D} e H H H L
    \} \,.
\end{split}
\end{align}
The $\Delta B = 1 = -\Delta L$ operators induce proton decay while the $\Delta L =2$ operators generically induce Majorana neutrino masses; neither give pure charged LFV. The operator $\bar{e}LLLH$ allows for LFV by three units, an  interesting experimental signature unsuppressed by neutrino mass~\cite{Heeck:2025jfs}.

\begin{figure*}[tb]
    \centering

    \begin{subfigure}[b]{0.4\columnwidth}
        \includegraphics[width=\textwidth]{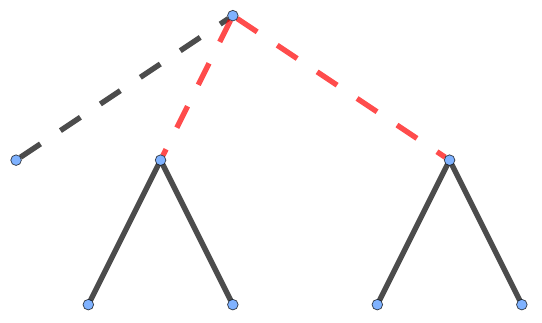}
        \caption{}
        \label{Topo_7_1}
    \end{subfigure}
    \hfill
    \begin{subfigure}[b]{0.4\columnwidth}
        \includegraphics[width=\textwidth]{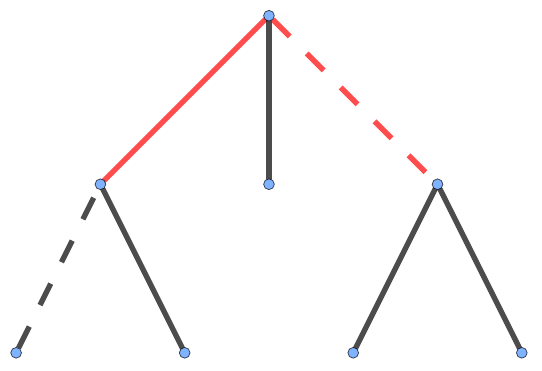}
        \caption{}
        \label{Topo_7_2}
    \end{subfigure}
    \hfill
    \begin{subfigure}[b]{0.4\columnwidth}
        \includegraphics[width=\textwidth]{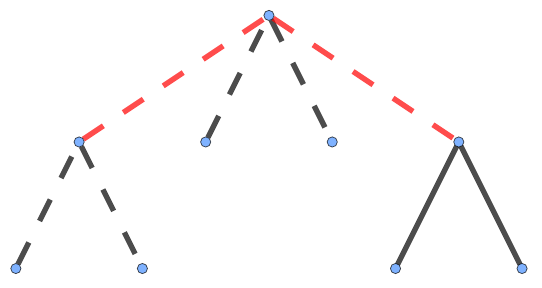}
        \caption{}
        \label{Topo_7_3}
    \end{subfigure}
    \hfill
    \begin{subfigure}[b]{0.4\columnwidth}
        \includegraphics[width=\textwidth]{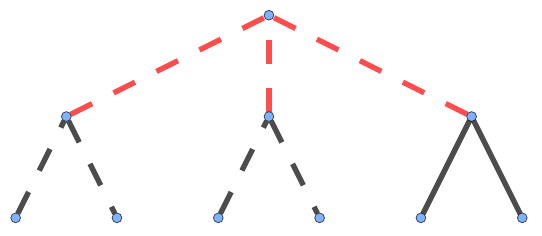}
        \caption{}
        \label{Topo_7_4}
    \end{subfigure}

    \begin{subfigure}[b]{0.4\columnwidth}
        \includegraphics[width=\textwidth]{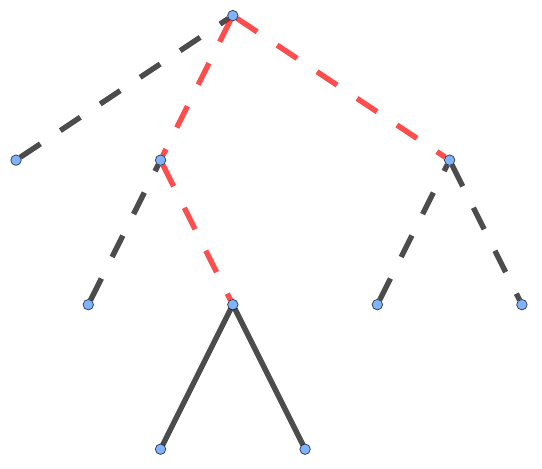}
        \caption{}
        \label{Topo_7_5}
    \end{subfigure}
    \hfill
    \begin{subfigure}[b]{0.4\columnwidth}
        \includegraphics[width=\textwidth]{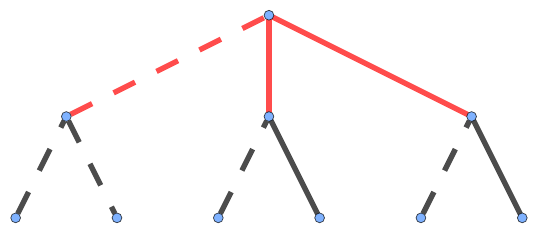}
        \caption{}
        \label{Topo_7_6}
    \end{subfigure}
    \hfill
    \begin{subfigure}[b]{0.4\columnwidth}
        \includegraphics[width=\textwidth]{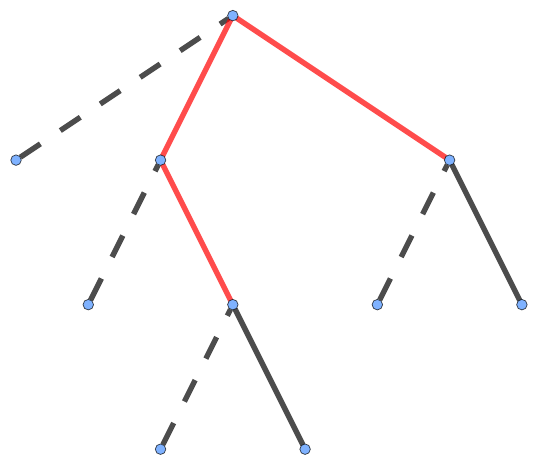}
        \caption{}
        \label{Topo_7_7}
    \end{subfigure}
    \hfill
    \begin{subfigure}[b]{0.4\columnwidth}
        \includegraphics[width=\textwidth]{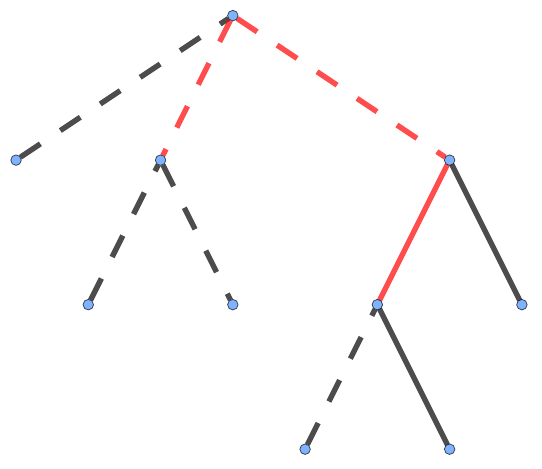}
        \caption{}
        \label{Topo_7_8}
    \end{subfigure}

    \begin{subfigure}[b]{0.4\columnwidth}
        \includegraphics[width=\textwidth]{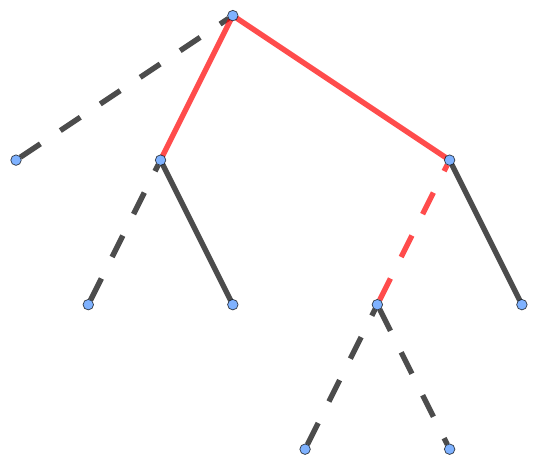}
        \caption{}
        \label{Topo_7_9}
    \end{subfigure}
    \hfill
    \begin{subfigure}[b]{0.4\columnwidth}
        \includegraphics[width=\textwidth]{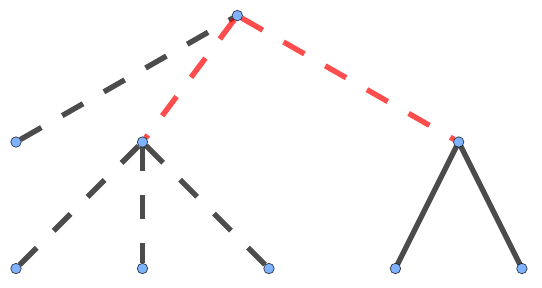}
        \caption{}
        \label{Topo_7_10}
    \end{subfigure}
    \hfill
    \begin{subfigure}[b]{0.4\columnwidth}
        \includegraphics[width=\textwidth]{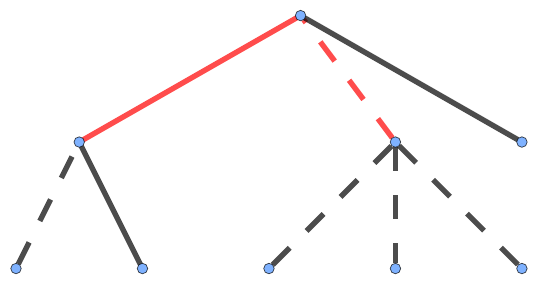}
        \caption{}
        \label{Topo_7_11}
    \end{subfigure}
    \hfill
    \begin{subfigure}[b]{0.4\columnwidth}
        \includegraphics[width=\textwidth]{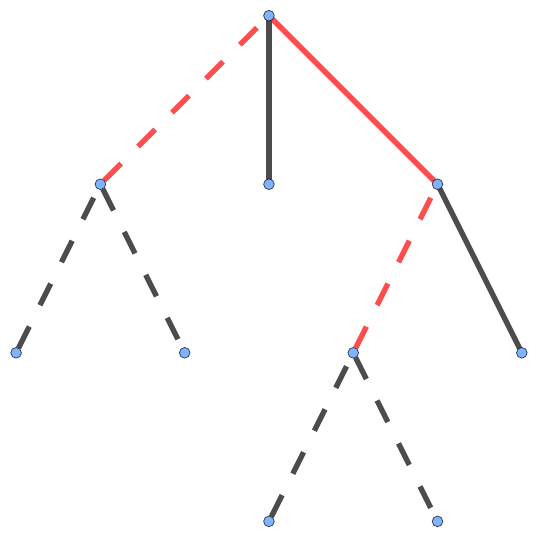}
        \caption{}
        \label{Topo_7_12}
    \end{subfigure}

    \caption{Topologies for tree-level UV completions of $d=7$ LFV operators.}
    \label{fig:dim7_topos}
\end{figure*}

Comprehensive UV completions for these operators can be found in Ref.~\cite{Li:2022abx}. Using the topologies from Fig.~\ref{fig:dim7_topos} and the code from Refs.~\cite{Heeck:2026dmh,Heeck:2026lhd}, we generated our own UV completions, listed in Tabs.~\ref{tab:d_7_1}--\ref{tab:d_7_3} for non-derivative and in Tab.~\ref{tab:d_7_der} for derivative operators. These agree with Ref.~\cite{Li:2022abx} for the most part, although we find two additional UV completions for the operator $HHH\bar{H}LL$, the first two (blue) ones in Tab.~\ref{tab:d_7_2}. These are also listed in Ref.~\cite{Bonnet:2009ej}. Aside from these omissions, Ref.~\cite{Li:2022abx} also redundantly lists some UV completions more than once.

\begin{table}[H]
    \centering
    \renewcommand{\arraystretch}{1.4} 
    \setlength{\tabcolsep}{10pt} 
    \begin{tabular}{|c|c c|c|}
        \hline
        Operator & \multicolumn{2}{c|}{UV Completion}   & Topo. \\
        \hline\hline
        $dd\bar{H}\bar{L}u$
        & $S_{(\mathbf{3}, \mathbf{1}, \tfrac{2}{3})}$ & $S_{(\mathbf{3}, \mathbf{2}, \tfrac{7}{6})}$ & Fig.~\ref{Topo_7_1}\\
        & $S_{(\mathbf{3}, \mathbf{1}, -\tfrac{1}{3})}$ & $S_{(\mathbf{3}, \mathbf{2}, \tfrac{1}{6})}$ & Fig.~\ref{Topo_7_1}\\
        & $F_{(\mathbf{3}, \mathbf{2}, \tfrac{1}{6})}$ & $S_{(\mathbf{3}, \mathbf{1}, \tfrac{2}{3})}$ & Fig.~\ref{Topo_7_2}\\
        & $F_{(\mathbf{3}, \mathbf{2}, -\tfrac{5}{6})}$ & $S_{(\mathbf{3}, \mathbf{1}, -\tfrac{1}{3})}$ & Fig.~\ref{Topo_7_2}\\
        & $F_{(\mathbf{1}, \mathbf{1}, 0)}$ & $S_{(\mathbf{3}, \mathbf{1}, \tfrac{2}{3})}$ & Fig.~\ref{Topo_7_2}\\
        & $F_{(\mathbf{3}, \mathbf{2}, -\tfrac{5}{6})}$ & $S_{(\mathbf{3}, \mathbf{2}, \tfrac{1}{6})}$ & Fig.~\ref{Topo_7_2}\\
        & $F_{(\mathbf{1}, \mathbf{1}, 0)}$ & $S_{(\mathbf{3}, \mathbf{1}, -\tfrac{1}{3})}$ & Fig.~\ref{Topo_7_2}\\
        & $F_{(\mathbf{3}, \mathbf{2}, \tfrac{1}{6})}$ & $S_{(\mathbf{3}, \mathbf{2}, \tfrac{1}{6})}$ & Fig.~\ref{Topo_7_2}\\
        & $F_{(\mathbf{3}, \mathbf{2}, -\tfrac{5}{6})}$ & $S_{(\mathbf{3}, \mathbf{2}, \tfrac{7}{6})}$ & Fig.~\ref{Topo_7_2}\\
        \hline
        $d\bar{H}\bar{L}QQ$
        & $S_{(\mathbf{3}, \mathbf{2}, \tfrac{1}{6})}$ & $S_{(\mathbf{3}, \mathbf{3}, -\tfrac{1}{3})}$ & Fig.~\ref{Topo_7_1}\\
        & $S_{(\mathbf{3}, \mathbf{1}, -\tfrac{1}{3})}$ & $S_{(\mathbf{3}, \mathbf{2}, \tfrac{1}{6})}$ & Fig.~\ref{Topo_7_1}\\
        & $F_{(\mathbf{3}, \mathbf{2}, -\tfrac{5}{6})}$ & $S_{(\mathbf{3}, \mathbf{3}, -\tfrac{1}{3})}$ & Fig.~\ref{Topo_7_2}\\
        & $F_{(\mathbf{3}, \mathbf{2}, -\tfrac{5}{6})}$ & $S_{(\mathbf{3}, \mathbf{1}, -\tfrac{1}{3})}$ & Fig.~\ref{Topo_7_2}\\
        & $F_{(\mathbf{1}, \mathbf{3}, 0)}$ & $S_{(\mathbf{3}, \mathbf{3}, -\tfrac{1}{3})}$ & Fig.~\ref{Topo_7_2}\\
        & $F_{(\mathbf{1}, \mathbf{1}, 0)}$ & $S_{(\mathbf{3}, \mathbf{1}, -\tfrac{1}{3})}$ & Fig.~\ref{Topo_7_2}\\
        & $F_{(\mathbf{3}, \mathbf{3}, -\tfrac{1}{3})}$ & $S_{(\mathbf{3}, \mathbf{2}, \tfrac{1}{6})}$ & Fig.~\ref{Topo_7_2}\\
        & $F_{(\mathbf{3}, \mathbf{1}, -\tfrac{1}{3})}$ & $S_{(\mathbf{3}, \mathbf{2}, \tfrac{1}{6})}$ & Fig.~\ref{Topo_7_2}\\
        \hline
        $dd\bar{e}\bar{H}Q$
        & $S_{(\mathbf{3}, \mathbf{1}, \tfrac{2}{3})}$ & $S_{(\mathbf{3}, \mathbf{2}, \tfrac{7}{6})}$ & Fig.~\ref{Topo_7_1}\\
        & $F_{(\mathbf{3}, \mathbf{1}, -\tfrac{1}{3})}$ & $S_{(\mathbf{3}, \mathbf{1}, \tfrac{2}{3})}$ & Fig.~\ref{Topo_7_2}\\
        & $F_{(\mathbf{3}, \mathbf{2}, -\tfrac{5}{6})}$ & $S_{(\mathbf{3}, \mathbf{2}, \tfrac{7}{6})}$ & Fig.~\ref{Topo_7_2}\\
        & $F_{(\mathbf{1}, \mathbf{2}, \tfrac{1}{2})}$ & $S_{(\mathbf{3}, \mathbf{1}, \tfrac{2}{3})}$ & Fig.~\ref{Topo_7_2}\\
        \hline
        $HLL\bar{Q}u$
        & $S_{(\mathbf{1}, \mathbf{2}, \tfrac{1}{2})}$ & $S_{(\mathbf{1}, \mathbf{3}, 1)}$ & Fig.~\ref{Topo_7_1}\\
        & $S_{(\mathbf{1}, \mathbf{1}, 1)}$ & $S_{(\mathbf{1}, \mathbf{2}, \tfrac{1}{2})}$ & Fig.~\ref{Topo_7_1}\\
        & $F_{(\mathbf{3}, \mathbf{2}, \tfrac{7}{6})}$ & $S_{(\mathbf{1}, \mathbf{3}, 1)}$ & Fig.~\ref{Topo_7_2}\\
        & $F_{(\mathbf{3}, \mathbf{2}, \tfrac{7}{6})}$ & $S_{(\mathbf{1}, \mathbf{1}, 1)}$ & Fig.~\ref{Topo_7_2}\\
        & $F_{(\mathbf{1}, \mathbf{3}, 0)}$ & $S_{(\mathbf{1}, \mathbf{2}, \tfrac{1}{2})}$ & Fig.~\ref{Topo_7_2}\\
        & $F_{(\mathbf{1}, \mathbf{1}, 0)}$ & $S_{(\mathbf{1}, \mathbf{2}, \tfrac{1}{2})}$ & Fig.~\ref{Topo_7_2}\\
        & $F_{(\mathbf{3}, \mathbf{3}, -\tfrac{1}{3})}$ & $S_{(\mathbf{1}, \mathbf{3}, 1)}$ & Fig.~\ref{Topo_7_2}\\
        & $F_{(\mathbf{3}, \mathbf{1}, -\tfrac{1}{3})}$ & $S_{(\mathbf{1}, \mathbf{1}, 1)}$ & Fig.~\ref{Topo_7_2}\\
        \hline
        $\bar{d}eHLu$
        & $S_{(\mathbf{3}, \mathbf{1}, -\tfrac{1}{3})}$ & $S_{(\mathbf{3}, \mathbf{2}, \tfrac{1}{6})}$ & Fig.~\ref{Topo_7_1}\\
        & $F_{(\mathbf{1}, \mathbf{1}, 0)}$ & $S_{(\mathbf{3}, \mathbf{1}, -\tfrac{1}{3})}$ & Fig.~\ref{Topo_7_2}\\
        & $F_{(\mathbf{3}, \mathbf{2}, \tfrac{7}{6})}$ & $S_{(\mathbf{3}, \mathbf{2}, \tfrac{1}{6})}$ & Fig.~\ref{Topo_7_2}\\
        & $F_{(\mathbf{3}, \mathbf{2}, -\tfrac{5}{6})}$ & $S_{(\mathbf{3}, \mathbf{1}, -\tfrac{1}{3})}$ & Fig.~\ref{Topo_7_2}\\
        & $F_{(\mathbf{1}, \mathbf{2}, \tfrac{1}{2})}$ & $S_{(\mathbf{3}, \mathbf{2}, \tfrac{1}{6})}$ & Fig.~\ref{Topo_7_2}\\
        \hline
    \end{tabular}
    \caption{UV completions for non-derivative $d = 7$ LFV operators, part 1.}
    \label{tab:d_7_1}
\end{table}

\begin{table}[H]
    \centering
    \renewcommand{\arraystretch}{1.4} 
    \setlength{\tabcolsep}{3pt} 
    \begin{tabular}{|c|c c|c|}
        \hline
        Operator & \multicolumn{2}{c|}{UV Completion}   & Topo. \\
        \hline \hline
        $\bar{d}HLLQ$
        & $S_{(\mathbf{3}, \mathbf{2}, \tfrac{1}{6})}$ & $S_{(\mathbf{3}, \mathbf{3}, -\tfrac{1}{3})}$ & Fig.~\ref{Topo_7_1}\\
        & $S_{(\mathbf{3}, \mathbf{1}, -\tfrac{1}{3})}$ & $S_{(\mathbf{3}, \mathbf{2}, \tfrac{1}{6})}$ & Fig.~\ref{Topo_7_1}\\
        & $S_{(\mathbf{1}, \mathbf{2}, \tfrac{1}{2})}$ & $S_{(\mathbf{1}, \mathbf{3}, 1)}$ & Fig.~\ref{Topo_7_1}\\
        & $S_{(\mathbf{1}, \mathbf{1}, 1)}$ & $S_{(\mathbf{1}, \mathbf{2}, \tfrac{1}{2})}$ & Fig.~\ref{Topo_7_1}\\
        & $F_{(\mathbf{1}, \mathbf{3}, 0)}$ & $S_{(\mathbf{3}, \mathbf{3}, -\tfrac{1}{3})}$ & Fig.~\ref{Topo_7_2}\\
        & $F_{(\mathbf{1}, \mathbf{1}, 0)}$ & $S_{(\mathbf{3}, \mathbf{1}, -\tfrac{1}{3})}$ & Fig.~\ref{Topo_7_2}\\
        & $F_{(\mathbf{3}, \mathbf{3}, \tfrac{2}{3})}$ & $S_{(\mathbf{1}, \mathbf{3}, 1)}$ & Fig.~\ref{Topo_7_2}\\
        & $F_{(\mathbf{3}, \mathbf{1}, \tfrac{2}{3})}$ & $S_{(\mathbf{1}, \mathbf{1}, 1)}$ & Fig.~\ref{Topo_7_2}\\
        & $F_{(\mathbf{3}, \mathbf{2}, -\tfrac{5}{6})}$ & $S_{(\mathbf{3}, \mathbf{3}, -\tfrac{1}{3})}$ & Fig.~\ref{Topo_7_2}\\
        & $F_{(\mathbf{3}, \mathbf{2}, -\tfrac{5}{6})}$ & $S_{(\mathbf{3}, \mathbf{1}, -\tfrac{1}{3})}$ & Fig.~\ref{Topo_7_2}\\
        & $F_{(\mathbf{1}, \mathbf{3}, 0)}$ & $S_{(\mathbf{1}, \mathbf{2}, \tfrac{1}{2})}$ & Fig.~\ref{Topo_7_2}\\
        & $F_{(\mathbf{1}, \mathbf{1}, 0)}$ & $S_{(\mathbf{1}, \mathbf{2}, \tfrac{1}{2})}$ & Fig.~\ref{Topo_7_2}\\
        & $F_{(\mathbf{3}, \mathbf{3}, \tfrac{2}{3})}$ & $S_{(\mathbf{3}, \mathbf{2}, \tfrac{1}{6})}$ & Fig.~\ref{Topo_7_2}\\
        & $F_{(\mathbf{3}, \mathbf{1}, \tfrac{2}{3})}$ & $S_{(\mathbf{3}, \mathbf{2}, \tfrac{1}{6})}$ & Fig.~\ref{Topo_7_2}\\
        & $F_{(\mathbf{3}, \mathbf{2}, -\tfrac{5}{6})}$ & $S_{(\mathbf{1}, \mathbf{3}, 1)}$ & Fig.~\ref{Topo_7_2}\\
        & $F_{(\mathbf{3}, \mathbf{2}, -\tfrac{5}{6})}$ & $S_{(\mathbf{1}, \mathbf{1}, 1)}$ & Fig.~\ref{Topo_7_2}\\
        & $F_{(\mathbf{1}, \mathbf{3}, 0)}$ & $S_{(\mathbf{3}, \mathbf{2}, \tfrac{1}{6})}$ & Fig.~\ref{Topo_7_2}\\
        & $F_{(\mathbf{1}, \mathbf{1}, 0)}$ & $S_{(\mathbf{3}, \mathbf{2}, \tfrac{1}{6})}$ & Fig.~\ref{Topo_7_2}\\
        \hline
        $dddH\bar{L}$
        & $S_{(\mathbf{3}, \mathbf{1}, \tfrac{2}{3})}$ & $S_{(\mathbf{3}, \mathbf{2}, \tfrac{1}{6})}$ & Fig.~\ref{Topo_7_1}\\
        & $F_{(\mathbf{3}, \mathbf{2}, \tfrac{1}{6})}$ & $S_{(\mathbf{3}, \mathbf{1}, \tfrac{2}{3})}$ & Fig.~\ref{Topo_7_2}\\
        & $F_{(\mathbf{1}, \mathbf{1}, 1)}$ & $S_{(\mathbf{3}, \mathbf{1}, \tfrac{2}{3})}$ & Fig.~\ref{Topo_7_2}\\
        & $F_{(\mathbf{3}, \mathbf{2}, \tfrac{1}{6})}$ & $S_{(\mathbf{3}, \mathbf{2}, \tfrac{1}{6})}$ & Fig.~\ref{Topo_7_2}\\
        \hline
        $\bar{e}HLLL$
        & $S_{(\mathbf{1}, \mathbf{2}, \tfrac{1}{2})}$ & $S_{(\mathbf{1}, \mathbf{3}, 1)}$ & Fig.~\ref{Topo_7_1}\\
        & $S_{(\mathbf{1}, \mathbf{1}, 1)}$ & $S_{(\mathbf{1}, \mathbf{2}, \tfrac{1}{2})}$ & Fig.~\ref{Topo_7_1}\\
        & $F_{(\mathbf{1}, \mathbf{3}, 0)}$ & $S_{(\mathbf{1}, \mathbf{3}, 1)}$ & Fig.~\ref{Topo_7_2}\\
        & $F_{(\mathbf{1}, \mathbf{1}, 0)}$ & $S_{(\mathbf{1}, \mathbf{1}, 1)}$ & Fig.~\ref{Topo_7_2}\\
        & $F_{(\mathbf{1}, \mathbf{2}, \tfrac{3}{2})}$ & $S_{(\mathbf{1}, \mathbf{3}, 1)}$ & Fig.~\ref{Topo_7_2}\\
        & $F_{(\mathbf{1}, \mathbf{2}, \tfrac{3}{2})}$ & $S_{(\mathbf{1}, \mathbf{1}, 1)}$ & Fig.~\ref{Topo_7_2}\\
        & $F_{(\mathbf{1}, \mathbf{3}, 0)}$ & $S_{(\mathbf{1}, \mathbf{2}, \tfrac{1}{2})}$ & Fig.~\ref{Topo_7_2}\\
        & $F_{(\mathbf{1}, \mathbf{1}, 0)}$ & $S_{(\mathbf{1}, \mathbf{2}, \tfrac{1}{2})}$ & Fig.~\ref{Topo_7_2}\\
        \hline
        $HHH\bar{H}LL$
        & {\color{blue}$S_{(\mathbf{1}, \mathbf{1}, 0)}$} & {\color{blue}$S_{(\mathbf{1}, \mathbf{3}, 1)}$} & Fig.~\ref{Topo_7_3}\\
        & {\color{blue}$S_{(\mathbf{1}, \mathbf{3}, 1)}$} & {\color{blue}$S_{(\mathbf{1}, \mathbf{3}, 1)}$} & Fig.~\ref{Topo_7_3}\\
        & $S_{(\mathbf{1}, \mathbf{3}, 1)}$ & $S_{(\mathbf{1}, \mathbf{4}, \tfrac{3}{2})}$ & Fig.~\ref{Topo_7_10}\\
        & $S_{(\mathbf{1}, \mathbf{3}, 1)}$ & $S_{(\mathbf{1}, \mathbf{4}, \tfrac{1}{2})}$ & Fig.~\ref{Topo_7_10}\\
        & $S_{(\mathbf{1}, \mathbf{2}, \tfrac{1}{2})}$ & $S_{(\mathbf{1}, \mathbf{3}, 1)}$ & Fig.~\ref{Topo_7_10}\\
        & $F_{(\mathbf{1}, \mathbf{3}, 1)}$ & $S_{(\mathbf{1}, \mathbf{4}, \tfrac{3}{2})}$ & Fig.~\ref{Topo_7_11}\\
        & $F_{(\mathbf{1}, \mathbf{3}, 0)}$ & $S_{(\mathbf{1}, \mathbf{4}, \tfrac{1}{2})}$ & Fig.~\ref{Topo_7_11}\\
        & $F_{(\mathbf{1}, \mathbf{3}, 0)}$ & $S_{(\mathbf{1}, \mathbf{2}, \tfrac{1}{2})}$ & Fig.~\ref{Topo_7_11}\\
        & $F_{(\mathbf{1}, \mathbf{1}, 0)}$ & $S_{(\mathbf{1}, \mathbf{2}, \tfrac{1}{2})}$ & Fig.~\ref{Topo_7_11}\\
        \hline
    \end{tabular}
    \caption{$d = 7$ UV completions, part 2.}
    \label{tab:d_7_2}
\end{table}

\begin{table}[H]
    \centering
    \renewcommand{\arraystretch}{1.2} 
    \setlength{\tabcolsep}{3pt} 
    \begin{tabular}{|c|c c c|c|}
        \hline
        Operator & \multicolumn{3}{c|}{UV Completion} & Topo. \\
        \hline\hline
        $HHH\bar{H}LL$
        & $F_{(\mathbf{1}, \mathbf{1}, 0)}$ & $F_{(\mathbf{1}, \mathbf{3}, 0)}$ & $S_{(\mathbf{1}, \mathbf{3}, 0)}$ & Fig.~\ref{Topo_7_6}\\
        & $F_{(\mathbf{1}, \mathbf{3}, 0)}$ & $F_{(\mathbf{1}, \mathbf{3}, 0)}$ & $S_{(\mathbf{1}, \mathbf{1}, 0)}$ & Fig.~\ref{Topo_7_6}\\
        & $F_{(\mathbf{1}, \mathbf{1}, 0)}$ & $F_{(\mathbf{1}, \mathbf{1}, 0)}$ & $S_{(\mathbf{1}, \mathbf{1}, 0)}$ & Fig.~\ref{Topo_7_6}\\
        & $F_{(\mathbf{1}, \mathbf{3}, 0)}$ & $F_{(\mathbf{1}, \mathbf{3}, 1)}$ & $S_{(\mathbf{1}, \mathbf{3}, 1)}$ & Fig.~\ref{Topo_7_6}\\
        & $F_{(\mathbf{1}, \mathbf{1}, 0)}$ & $F_{(\mathbf{1}, \mathbf{3}, 1)}$ & $S_{(\mathbf{1}, \mathbf{3}, 1)}$ & Fig.~\ref{Topo_7_6}\\
        & $F_{(\mathbf{1}, \mathbf{3}, 0)}$ & $F_{(\mathbf{1}, \mathbf{3}, 0)}$ & $F_{(\mathbf{1}, \mathbf{4}, \tfrac{1}{2})}$ & Fig.~\ref{Topo_7_7}\\
        & $F_{(\mathbf{1}, \mathbf{2}, \tfrac{1}{2})}$ & $F_{(\mathbf{1}, \mathbf{3}, 0)}$ & $F_{(\mathbf{1}, \mathbf{3}, 0)}$ & Fig.~\ref{Topo_7_7}\\
        & $F_{(\mathbf{1}, \mathbf{1}, 0)}$ & $F_{(\mathbf{1}, \mathbf{2}, \tfrac{1}{2})}$ & $F_{(\mathbf{1}, \mathbf{3}, 0)}$ & Fig.~\ref{Topo_7_7}\\
        & $F_{(\mathbf{1}, \mathbf{1}, 0)}$ & $F_{(\mathbf{1}, \mathbf{1}, 0)}$ & $F_{(\mathbf{1}, \mathbf{2}, \tfrac{1}{2})}$ & Fig.~\ref{Topo_7_7}\\
        & $F_{(\mathbf{1}, \mathbf{3}, 0)}$ & $F_{(\mathbf{1}, \mathbf{3}, 1)}$ & $F_{(\mathbf{1}, \mathbf{4}, \tfrac{1}{2})}$ & Fig.~\ref{Topo_7_7}\\
        & $F_{(\mathbf{1}, \mathbf{2}, \tfrac{1}{2})}$ & $F_{(\mathbf{1}, \mathbf{3}, 0)}$ & $F_{(\mathbf{1}, \mathbf{3}, 1)}$ & Fig.~\ref{Topo_7_7}\\
        & $F_{(\mathbf{1}, \mathbf{1}, 0)}$ & $F_{(\mathbf{1}, \mathbf{2}, \tfrac{1}{2})}$ & $F_{(\mathbf{1}, \mathbf{3}, 1)}$ & Fig.~\ref{Topo_7_7}\\
        & $F_{(\mathbf{1}, \mathbf{3}, 0)}$ & $S_{(\mathbf{1}, \mathbf{3}, 1)}$ & $S_{(\mathbf{1}, \mathbf{4}, \tfrac{1}{2})}$ & Fig.~\ref{Topo_7_8}\\
        & $F_{(\mathbf{1}, \mathbf{3}, 0)}$ & $S_{(\mathbf{1}, \mathbf{2}, \tfrac{1}{2})}$ & $S_{(\mathbf{1}, \mathbf{3}, 1)}$ & Fig.~\ref{Topo_7_8}\\
        & $F_{(\mathbf{1}, \mathbf{1}, 0)}$ & $S_{(\mathbf{1}, \mathbf{2}, \tfrac{1}{2})}$ & $S_{(\mathbf{1}, \mathbf{3}, 1)}$ & Fig.~\ref{Topo_7_8}\\
        & $F_{(\mathbf{1}, \mathbf{3}, 0)}$ & $S_{(\mathbf{1}, \mathbf{3}, 0)}$ & $S_{(\mathbf{1}, \mathbf{4}, \tfrac{1}{2})}$ & Fig.~\ref{Topo_7_8}\\
        & $F_{(\mathbf{1}, \mathbf{3}, 0)}$ & $S_{(\mathbf{1}, \mathbf{2}, \tfrac{1}{2})}$ & $S_{(\mathbf{1}, \mathbf{3}, 0)}$ & Fig.~\ref{Topo_7_8}\\
        & $F_{(\mathbf{1}, \mathbf{3}, 0)}$ & $S_{(\mathbf{1}, \mathbf{1}, 0)}$ & $S_{(\mathbf{1}, \mathbf{2}, \tfrac{1}{2})}$ & Fig.~\ref{Topo_7_8}\\
        & $F_{(\mathbf{1}, \mathbf{1}, 0)}$ & $S_{(\mathbf{1}, \mathbf{2}, \tfrac{1}{2})}$ & $S_{(\mathbf{1}, \mathbf{3}, 0)}$ & Fig.~\ref{Topo_7_8}\\
        & $F_{(\mathbf{1}, \mathbf{1}, 0)}$ & $S_{(\mathbf{1}, \mathbf{1}, 0)}$ & $S_{(\mathbf{1}, \mathbf{2}, \tfrac{1}{2})}$ & Fig.~\ref{Topo_7_8}\\
        & $F_{(\mathbf{1}, \mathbf{3}, 1)}$ & $S_{(\mathbf{1}, \mathbf{3}, 1)}$ & $S_{(\mathbf{1}, \mathbf{4}, \tfrac{3}{2})}$ & Fig.~\ref{Topo_7_8}\\
        & $F_{(\mathbf{1}, \mathbf{3}, 0)}$ & $F_{(\mathbf{1}, \mathbf{4}, \tfrac{1}{2})}$ & $S_{(\mathbf{1}, \mathbf{3}, 1)}$ & Fig.~\ref{Topo_7_9}\\
        & $F_{(\mathbf{1}, \mathbf{2}, \tfrac{1}{2})}$ & $F_{(\mathbf{1}, \mathbf{3}, 0)}$ & $S_{(\mathbf{1}, \mathbf{3}, 1)}$ & Fig.~\ref{Topo_7_9}\\
        & $F_{(\mathbf{1}, \mathbf{1}, 0)}$ & $F_{(\mathbf{1}, \mathbf{2}, \tfrac{1}{2})}$ & $S_{(\mathbf{1}, \mathbf{3}, 1)}$ & Fig.~\ref{Topo_7_9}\\
        & $F_{(\mathbf{1}, \mathbf{3}, 1)}$ & $F_{(\mathbf{1}, \mathbf{4}, \tfrac{1}{2})}$ & $S_{(\mathbf{1}, \mathbf{3}, 1)}$ & Fig.~\ref{Topo_7_9}\\
        & $F_{(\mathbf{1}, \mathbf{2}, \tfrac{1}{2})}$ & $F_{(\mathbf{1}, \mathbf{3}, 1)}$ & $S_{(\mathbf{1}, \mathbf{3}, 1)}$ & Fig.~\ref{Topo_7_9}\\
        & $F_{(\mathbf{1}, \mathbf{3}, 0)}$ & $F_{(\mathbf{1}, \mathbf{4}, \tfrac{1}{2})}$ & $S_{(\mathbf{1}, \mathbf{3}, 0)}$ & Fig.~\ref{Topo_7_9}\\
        & $F_{(\mathbf{1}, \mathbf{2}, \tfrac{1}{2})}$ & $F_{(\mathbf{1}, \mathbf{3}, 0)}$ & $S_{(\mathbf{1}, \mathbf{3}, 0)}$ & Fig.~\ref{Topo_7_9}\\
        & $F_{(\mathbf{1}, \mathbf{1}, 0)}$ & $F_{(\mathbf{1}, \mathbf{2}, \tfrac{1}{2})}$ & $S_{(\mathbf{1}, \mathbf{3}, 0)}$ & Fig.~\ref{Topo_7_9}\\
        & $F_{(\mathbf{1}, \mathbf{2}, \tfrac{1}{2})}$ & $F_{(\mathbf{1}, \mathbf{3}, 0)}$ & $S_{(\mathbf{1}, \mathbf{1}, 0)}$ & Fig.~\ref{Topo_7_9}\\
        & $F_{(\mathbf{1}, \mathbf{1}, 0)}$ & $F_{(\mathbf{1}, \mathbf{2}, \tfrac{1}{2})}$ & $S_{(\mathbf{1}, \mathbf{1}, 0)}$ & Fig.~\ref{Topo_7_9}\\
        & $F_{(\mathbf{1}, \mathbf{4}, \tfrac{1}{2})}$ & $S_{(\mathbf{1}, \mathbf{3}, 0)}$ & $S_{(\mathbf{1}, \mathbf{3}, 1)}$ & Fig.~\ref{Topo_7_12}\\
        & $F_{(\mathbf{1}, \mathbf{2}, \tfrac{1}{2})}$ & $S_{(\mathbf{1}, \mathbf{3}, 0)}$ & $S_{(\mathbf{1}, \mathbf{3}, 1)}$ & Fig.~\ref{Topo_7_12}\\
        & $F_{(\mathbf{1}, \mathbf{2}, \tfrac{1}{2})}$ & $S_{(\mathbf{1}, \mathbf{1}, 0)}$ & $S_{(\mathbf{1}, \mathbf{3}, 1)}$ & Fig.~\ref{Topo_7_12}\\
        & $S_{(\mathbf{1}, \mathbf{3}, 0)}$ & $S_{(\mathbf{1}, \mathbf{3}, 1)}$ & $S_{(\mathbf{1}, \mathbf{3}, 1)}$ & Fig.~\ref{Topo_7_4}\\
        & $S_{(\mathbf{1}, \mathbf{1}, 0)}$ & $S_{(\mathbf{1}, \mathbf{3}, 1)}$ & $S_{(\mathbf{1}, \mathbf{3}, 1)}$ & Fig.~\ref{Topo_7_4}\\
        & $S_{(\mathbf{1}, \mathbf{3}, 1)}$ & $S_{(\mathbf{1}, \mathbf{3}, 1)}$ & $S_{(\mathbf{1}, \mathbf{4}, \tfrac{1}{2})}$ & Fig.~\ref{Topo_7_5}\\
        & $S_{(\mathbf{1}, \mathbf{2}, \tfrac{1}{2})}$ & $S_{(\mathbf{1}, \mathbf{3}, 1)}$ & $S_{(\mathbf{1}, \mathbf{3}, 1)}$ & Fig.~\ref{Topo_7_5}\\
        & $S_{(\mathbf{1}, \mathbf{3}, 1)}$ & $S_{(\mathbf{1}, \mathbf{3}, 1)}$ & $S_{(\mathbf{1}, \mathbf{4}, \tfrac{3}{2})}$ & Fig.~\ref{Topo_7_5}\\
        & $S_{(\mathbf{1}, \mathbf{3}, 0)}$ & $S_{(\mathbf{1}, \mathbf{3}, 1)}$ & $S_{(\mathbf{1}, \mathbf{4}, \tfrac{1}{2})}$ & Fig.~\ref{Topo_7_5}\\
        & $S_{(\mathbf{1}, \mathbf{2}, \tfrac{1}{2})}$ & $S_{(\mathbf{1}, \mathbf{3}, 0)}$ & $S_{(\mathbf{1}, \mathbf{3}, 1)}$ & Fig.~\ref{Topo_7_5}\\
        & $S_{(\mathbf{1}, \mathbf{1}, 0)}$ & $S_{(\mathbf{1}, \mathbf{2}, \tfrac{1}{2})}$ & $S_{(\mathbf{1}, \mathbf{3}, 1)}$ & Fig.~\ref{Topo_7_5}\\
        \hline
    \end{tabular}
    \caption{$d = 7$ UV completions, part 3.}
    \label{tab:d_7_3}
\end{table}

\begin{table}[H]
    \centering
    \renewcommand{\arraystretch}{1.4} 
    \setlength{\tabcolsep}{10pt} 
    \begin{tabular}{|c|c c|c|}
        \hline
        Operator & \multicolumn{2}{c|}{UV Completion}   & Topo. \\
        \hline        \hline
        $\mathcal{D} \mathcal{D} H H L L$
        & $S_{(\mathbf{1}, \mathbf{3}, 1)}$ & - & Fig.~\ref{Topo_5_1} \\
        & $F_{(\mathbf{1}, \mathbf{3}, 0)}$ & - & Fig.~\ref{Topo_5_2} \\
        & $F_{(\mathbf{1}, \mathbf{1}, 0)}$ & - & Fig.~\ref{Topo_5_2} \\
        \hline
        $\mathcal{D} e H H H L$
        & $F_{(\mathbf{1}, \mathbf{2}, \tfrac{1}{2})}$ & $S_{(\mathbf{1}, \mathbf{3}, 1)}$ & Fig.~\ref{Topo_6_3} \\
        & $F_{(\mathbf{1}, \mathbf{3}, 0)}$ & $S_{(\mathbf{1}, \mathbf{3}, 1)}$ & Fig.~\ref{Topo_6_3} \\
        & $F_{(\mathbf{1}, \mathbf{2}, \tfrac{1}{2})}$ & $F_{(\mathbf{1}, \mathbf{3}, 0)}$ & Fig.~\ref{Topo_6_2} \\
        & $F_{(\mathbf{1}, \mathbf{2}, \tfrac{1}{2})}$ & $F_{(\mathbf{1}, \mathbf{1}, 0)}$ & Fig.~\ref{Topo_6_2} \\
        \hline
    \end{tabular}
    \caption{Tree-level UV completions for derivative $d = 7$ LFV operators. 
    }
    \label{tab:d_7_der}
\end{table}

Notice that once again operators with field-strength tensors have no tree-level UV completions, nor do the operators containing one derivative and four fermions, so they are not listed in the tables. The baryon-number-violating ones,  $\mathcal{D} d d Q \bar{L} $ and  $\mathcal{D} d d d \bar{e} $, instead arise at loop level from $d=9$ operators, or at loop level by stitching together the $\Delta L = 2$ Weinberg operator and the $\Delta B = \Delta L = 1$ dimension-6 operators.
The operator $\mathcal{D} \mathcal{D} H H L L$ from Tab.~\ref{tab:d_7_der} simply arises from the seesaw mechanisms of Tab.~\ref{tab:d_5} once the mediator propagator is expanded to next order in momentum; as a result, they are always subdominant to $HHLL$ and of little practical interest.
Lastly, the UV completions for $\mathcal{D} e H H H L$ in Tab.~\ref{tab:d_7_der} all contain mediators that by themselves induce the Weinberg operator, generically suppressing this operator as well.

Of some interest are the models in Tab.~\ref{tab:d_7_Higgs_asymmetry}: they all form valid Feynman diagrams, which however vanish if evaluated for constant propagators since they are antisymmetric under $H\leftrightarrow H$ exchange. These UV completions therefore do not map onto $d=7$ SMEFT operators, but rather the $d=9$ operators $\mathcal{D}\mathcal{D}HHH\bar{H}LL$ upon expanding the propagators to second order in the large masses~\cite{Gargalionis:2020xvt}. The derivatives ensure that this operator does not induce neutrino masses after electroweak symmetry breaking, despite being a $\Delta L = 2$ operator. Still,  Majorana neutrino masses are generated at loop level, so the models of Tab.~\ref{tab:d_7_Higgs_asymmetry} can be viewed as radiative-neutrino-mass models, some of them being discussed in Refs.~\cite{Law:2013dya,Gargalionis:2020xvt}. Some of the models  in Tab.~\ref{tab:d_7_Higgs_asymmetry} contain particles that by themselves generate the Weinberg operator, i.e.~can be found in Tab.~\ref{tab:d_5}, but others, e.g.~the two scalars $S_{(\mathbf{1}, \mathbf{1}, 1)}$ and $S_{(\mathbf{1}, \mathbf{3}, 0)}$, lead to genuine loop neutrino masses, recently discussed in exactly this context in  Ref.~\cite{Esser:2026ehc}. 

\begin{table}[H]
    \centering
    \renewcommand{\arraystretch}{1.4} 
    \setlength{\tabcolsep}{4pt} 
    \begin{tabular}{|c|c c c|c|}
        \hline
        Operator & \multicolumn{3}{c|}{UV Completion} & Topo. \\
        \hline\hline
        $HHH\bar{H}LL$
        & $S_{(\mathbf{1}, \mathbf{3}, 0)}$ & $S_{(\mathbf{1}, \mathbf{3}, 1)}$ & - & Fig.~\ref{Topo_7_3}\\
        & $S_{(\mathbf{1}, \mathbf{1}, 1)}$ & $S_{(\mathbf{1}, \mathbf{3}, 0)}$ & - & Fig.~\ref{Topo_7_3}\\
        & $S_{(\mathbf{1}, \mathbf{1}, 1)}$ & $S_{(\mathbf{1}, \mathbf{3}, 1)}$ & - & Fig.~\ref{Topo_7_3}\\
        & $S_{(\mathbf{1}, \mathbf{1}, 1)}$ & $S_{(\mathbf{1}, \mathbf{3}, 0)}$ & $S_{(\mathbf{1}, \mathbf{3}, 1)}$ & Fig.~\ref{Topo_7_4}\\
        & $S_{(\mathbf{1}, \mathbf{1}, 1)}$ & $S_{(\mathbf{1}, \mathbf{2}, \tfrac{1}{2})}$ & $S_{(\mathbf{1}, \mathbf{3}, 1)}$ & Fig.~\ref{Topo_7_5}\\
        & $S_{(\mathbf{1}, \mathbf{2}, \tfrac{3}{2})}$ & $S_{(\mathbf{1}, \mathbf{3}, 1)}$ & $S_{(\mathbf{1}, \mathbf{3}, 1)}$ & Fig.~\ref{Topo_7_5}\\
        & $S_{(\mathbf{1}, \mathbf{1}, 1)}$ & $S_{(\mathbf{1}, \mathbf{2}, \tfrac{3}{2})}$ & $S_{(\mathbf{1}, \mathbf{3}, 1)}$ & Fig.~\ref{Topo_7_5}\\
        & $S_{(\mathbf{1}, \mathbf{1}, 1)}$ & $S_{(\mathbf{1}, \mathbf{2}, \tfrac{1}{2})}$ & $S_{(\mathbf{1}, \mathbf{3}, 0)}$ & Fig.~\ref{Topo_7_5}\\
        & $S_{(\mathbf{1}, \mathbf{1}, 0)}$ & $S_{(\mathbf{1}, \mathbf{1}, 1)}$ & $S_{(\mathbf{1}, \mathbf{2}, \tfrac{1}{2})}$ & Fig.~\ref{Topo_7_5}\\
        & $F_{(\mathbf{1}, \mathbf{3}, 0)}$ & $F_{(\mathbf{1}, \mathbf{3}, 0)}$ & $S_{(\mathbf{1}, \mathbf{3}, 0)}$ & Fig.~\ref{Topo_7_6}\\
        & $F_{(\mathbf{1}, \mathbf{1}, 1)}$ & $F_{(\mathbf{1}, \mathbf{3}, 0)}$ & $S_{(\mathbf{1}, \mathbf{3}, 1)}$ & Fig.~\ref{Topo_7_6}\\
        & $F_{(\mathbf{1}, \mathbf{1}, 1)}$ & $F_{(\mathbf{1}, \mathbf{2}, \tfrac{1}{2})}$ & $F_{(\mathbf{1}, \mathbf{3}, 0)}$ & Fig.~\ref{Topo_7_7}\\
        & $F_{(\mathbf{1}, \mathbf{1}, 0)}$ & $F_{(\mathbf{1}, \mathbf{1}, 1)}$ & $F_{(\mathbf{1}, \mathbf{2}, \tfrac{1}{2})}$ & Fig.~\ref{Topo_7_7}\\
        & $F_{(\mathbf{1}, \mathbf{3}, 1)}$ & $S_{(\mathbf{1}, \mathbf{2}, \tfrac{3}{2})}$ & $S_{(\mathbf{1}, \mathbf{3}, 1)}$ & Fig.~\ref{Topo_7_8}\\
        & $F_{(\mathbf{1}, \mathbf{1}, 1)}$ & $S_{(\mathbf{1}, \mathbf{2}, \tfrac{3}{2})}$ & $S_{(\mathbf{1}, \mathbf{3}, 1)}$ & Fig.~\ref{Topo_7_8}\\
        & $F_{(\mathbf{1}, \mathbf{1}, 1)}$ & $F_{(\mathbf{1}, \mathbf{2}, \tfrac{1}{2})}$ & $S_{(\mathbf{1}, \mathbf{3}, 1)}$ & Fig.~\ref{Topo_7_9}\\
        & $S_{(\mathbf{1}, \mathbf{1}, 1)}$ & $S_{(\mathbf{1}, \mathbf{2}, \tfrac{1}{2})}$ & - & Fig.~\ref{Topo_7_10}\\
        \hline
    \end{tabular}
    \caption{Models that seemingly UV-complete $HHH\bar{H}LL$ but vanish due to Higgs asymmetry and actually generate  $\mathcal{D}\mathcal{D}HHH\bar{H}LL$ at $d=9$, see text for details.}
    \label{tab:d_7_Higgs_asymmetry}
\end{table}

\clearpage

\section{\texorpdfstring{$d = 8$}{d = 8}}
\label{sec:UV8}

\begin{table}[H]
    \centering
    \renewcommand{\arraystretch}{1.4} 
    \setlength{\tabcolsep}{5pt} 
    \begin{tabular}{|c|c|}
        \hline
        Compact & Explicit \\
        \hline \hline
        $\mathcal{O}_{LeH^5}$ 
        & $\left(H^\dagger H\right)^2 
        \left(\bar{L}_p e_r H\right)$
        \\ $\mathcal{O}^{(1)}_{LeQdH^2}$ 
        & $(\bar{L}^j_p e_r)
        (\bar{d}_s Q_{tj})
        (H^\dagger H)$
        \\ $\mathcal{O}^{(2)}_{LeQdH^2}$ 
        & $(\bar{L}_p e_r)\tau^I
        (\bar{d}_s Q_t)
        (H^\dagger \tau^I H)$        
        \\ $\mathcal{O}^{(3)}_{LeQdH^2}$ 
        & $(\bar{L}_p e_r H)
        (\bar{Q}_s d_t H)$ 
        \\ $\mathcal{O}^{(4)}_{LeQdH^2}$ 
        & $(\bar{L}_p \sigma_{\mu \nu} e_r H)
        (\bar{Q}_s \sigma^{\mu \nu} d_t H)$
        \\ $\mathcal{O}_{L^2udH^2}$ 
        & $(\bar{L}_p d_r H)
        (\tilde{H}^\dagger \bar{u}_s L_t)$
        \\ $\mathcal{O}^{(1)}_{LeQuH^2}$ 
        & $(\bar{L}^j_p e_r) 
        \epsilon_{jk} 
        (\bar{Q}^k_s u_t)
        (H^\dagger H)$ 
        \\ $\mathcal{O}^{(2)}_{LeQuH^2}$ 
        & $(\bar{L}^j_p e_r)
        (\tau^I \epsilon)_{jk} 
        (\bar{Q}^k_s u_t)
        (H^\dagger \tau^I H)$ 
        \\ $\mathcal{O}^{(3)}_{LeQuH^2}$ 
        & $(\bar{L}^j_p \sigma_{\mu \nu} e_r) 
        \epsilon_{jk} 
        (\bar{Q}^k_s \sigma^{\mu \nu} u_t)
        (H^\dagger H)$
        \\ $\mathcal{O}^{(4)}_{LeQuH^2}$ 
        & $(\bar{L}^j_p \sigma_{\mu \nu} e_r)
        (\tau^I \epsilon)_{jk} 
        (\bar{Q}^k_s \sigma^{\mu \nu} u_t)
        (H^\dagger \tau^I H)$
        \\ $\mathcal{O}^{(5)}_{LeQuH^2}$ 
        & $(\bar{L}_p e_r H)(\tilde{H}^\dagger \bar{u}_s Q_t)$ \\
        $\mathcal{O}^{(1)}_{LQudH^2}$ 
        & $\epsilon_{\alpha \beta \gamma} \epsilon_{jk} 
        \left(d^\alpha_p C u^\beta_r\right) 
        \left(Q^{j\gamma}_s C L^k_t\right)
        (H^\dagger H)$
        \\ $\mathcal{O}^{(2)}_{LQudH^2}$ 
        & $\epsilon_{\alpha \beta \gamma} (\epsilon \tau^I)_{jk} 
        \left(d^\alpha_p C u^\beta_r\right) 
        \left(Q^{j\gamma}_s C L^k_t\right)
        (H^\dagger \tau^I H)$
        \\ $\mathcal{O}_{eQ^2uH^2}$ 
        & $\epsilon_{\alpha \beta \gamma} \epsilon_{jk} 
        \left(Q^{j \alpha}_p C Q^{m \beta}_r\right) 
        \left(u^\gamma_s C e_t\right)
        (H^\dagger_m H^k)$
        \\$\mathcal{O}^{(1)}_{LQ^3H^2}$ 
        & $\epsilon_{\alpha \beta \gamma} \epsilon_{mn} \epsilon_{jk} 
        \left(Q^{m \alpha}_p C Q^{j \beta}_r\right) 
        \left(Q^{k \gamma}_s C L^n_t\right)
        (H^\dagger H)$
        \\ $\mathcal{O}^{(2)}_{LQ^3H^2}$ 
        & $\epsilon_{\alpha \beta \gamma} (\epsilon \tau^I)_{mn} \epsilon_{jk} 
        \left(Q^{m \alpha}_p C Q^{j \beta}_r\right) 
        \left(Q^{k \gamma}_s C L^n_t\right)
        (H^\dagger \tau^I H)$
        \\ $\mathcal{O}^{(3)}_{LQ^3H^2}$ 
        & $\epsilon_{\alpha \beta \gamma} \epsilon_{mn} (\epsilon \tau^I)_{jk} 
        \left(Q^{m \alpha}_p C Q^{j \beta}_r\right)
        \left(Q^{k \gamma}_s C L^n_t\right)
        (H^\dagger \tau^I H)$
        \\ $\mathcal{O}_{eu^2dH^2}$ 
        & $\epsilon_{\alpha \beta \gamma} 
        \left(d^\alpha_p C u^\beta_r\right)
        \left(u^\gamma_s C e_t\right)
        (H^\dagger H)$
        \\ $\mathcal{O}_{LQu^2H^2}$ 
        & $\epsilon_{\alpha \beta \gamma} \epsilon_{jk} \epsilon_{mn} 
        \left(L^j_p C Q^{m \alpha}_r\right)
        \left(u^\beta_s C u^\gamma_t\right)
        \tilde{H}^k \tilde{H}^n$ 
        \\ $\mathcal{O}_{LQd^2H^2}$ 
        & $\epsilon_{\alpha \beta \gamma} \epsilon_{jk} \epsilon_{mn} 
        \left(L^j_p Q^{m \alpha}_r\right)
        \left(d^\beta_s C d^\gamma_t\right) 
        H^k H^n$ 
        \\ $\mathcal{O}_{eQ^2dH^2}$ 
        & $\epsilon_{\alpha \beta \gamma} \epsilon_{jk} \epsilon_{mn} 
        \left(e_p d^\alpha_r\right)
        \left(Q^{j \beta}_s C Q^{m \gamma}_t\right) 
        H^k H^n$ 
        \\ $\mathcal{O}^{(1)}_{L^4H^2}$ 
        & $(\bar{L}_p \gamma^\mu L_r)
        (\bar{L}_s \gamma_\mu L_t)
        (H^\dagger H)$ 
        \\ $\mathcal{O}^{(2)}_{L^4H^2}$ 
        & $(\bar{L}_p \gamma^\mu L_r)
        (\bar{L}_s \gamma_\mu \tau^I L_t)
        (H^\dagger \tau^I H)$ 
        \\ $\mathcal{O}^{(1)}_{L^2Q^2H^2}$ 
        & $(\bar{L}_p \gamma^\mu L_r)
        (\bar{Q}_s \gamma_\mu Q_t)
        (H^\dagger H)$ 
        \\ $\mathcal{O}^{(2)}_{L^2Q^2H^2}$ 
        & $(\bar{L}_p \gamma^\mu \tau^I L_r)
        (\bar{Q}_s \gamma_\mu Q_t)
        (H^\dagger \tau^I H)$ 
        \\ $\mathcal{O}^{(3)}_{L^2Q^2H^2}$ & 
        $(\bar{L}_p \gamma^\mu \tau^I L_r)
        (\bar{Q}_s \gamma_\mu \tau^I Q_t)
        (H^\dagger H)$ 
        \\ $\mathcal{O}^{(4)}_{L^2Q^2H^2}$ 
        & $(\bar{L}_p \gamma^\mu L_r)
        (\bar{Q}_s \gamma_\mu \tau^I Q_t)
        (H^\dagger \tau^I H)$ 
        \\ $\mathcal{O}^{(5)}_{L^2Q^2H^2}$ 
        & $\epsilon^{IJK}
        (\bar{L}_p \gamma^\mu \tau^I L_r)
        (\bar{Q}_s \gamma_\mu \tau^J Q_t)
        (H^\dagger \tau^K H)$ 
        \\ $\mathcal{O}_{e^4H^2}$ 
        & $(\bar{e}_p \gamma^\mu e_r)
        (\bar{e}_s \gamma_\mu e_t)
        (H^\dagger H)$ 
        \\ $\mathcal{O}_{e^2u^2H^2}$ 
        & $(\bar{e}_p \gamma^\mu e_r)
        (\bar{u}_s \gamma_\mu u_t)
        (H^\dagger H)$ 
        \\ $\mathcal{O}_{e^2d^2H^2}$ 
        & $(\bar{e}_p \gamma^\mu e_r)
        (\bar{d}_s \gamma_\mu d_t)
        (H^\dagger H)$ 
        \\ $\mathcal{O}^{(1)}_{L^2e^2H^2}$ 
        & $(\bar{L}_p \gamma^\mu L_r)
        (\bar{e}_s \gamma_\mu e_t)
        (H^\dagger H)$ 
        \\ $\mathcal{O}^{(2)}_{L^2e^2H^2}$ 
        & $(\bar{L}_p \gamma^\mu \tau^I L_r)
        (\bar{e}_s \gamma_\mu e_t)
        (H^\dagger \tau^I H)$
        \\ $\mathcal{O}^{(3)}_{L^2e^2H^2}$ 
        & $(\bar{L}_p e_r H)
        (\bar{L}_s e_t H)$
        \\ $\mathcal{O}^{(1)}_{L^2u^2H^2}$ 
        & $(\bar{L}_p \gamma^\mu L_r)
        (\bar{u}_s \gamma_\mu u_t)
        (H^\dagger H)$ 
        \\ $\mathcal{O}^{(2)}_{L^2u^2H^2}$ 
        & $(\bar{L}_p \gamma^\mu \tau^I L_r)
        (\bar{u}_s \gamma_\mu u_t)
        (H^\dagger \tau^I H)$ 
        \\ $\mathcal{O}^{(1)}_{L^2d^2H^2}$ 
        & $(\bar{L}_p \gamma^\mu L_r)
        (\bar{d}_s \gamma_\mu d_t)
        (H^\dagger H)$ 
        \\ $\mathcal{O}^{(2)}_{L^2d^2H^2}$ 
        & $(\bar{L}_p \gamma^\mu \tau^I L_r)
        (\bar{d}_s \gamma_\mu d_t)
        (H^\dagger \tau^I H)$ 
        \\ $\mathcal{O}^{(1)}_{Q^2e^2H^2}$ 
        & $(\bar{Q}_p \gamma^\mu Q_r)
        (\bar{e}_s \gamma_\mu e_t)
        (H^\dagger H)$ 
        \\ $\mathcal{O}^{(2)}_{Q^2e^2H^2}$ 
        & $(\bar{Q}_p \gamma^\mu \tau^I Q_r)
        (\bar{e}_s \gamma_\mu e_t)
        (H^\dagger \tau^I H)$ 
        \\\hline
    \end{tabular}
    \caption{Murphy's basis for non-derivative $d=8$ LFV SMEFT operators with explicit index contractions~\cite{Murphy:2020rsh}.}
    \label{tab:Murphy_basis}
\end{table}

Finally, we come to the $d=8$ LFV SMEFT operators. The quantum numbers are identical to $d=6$, i.e.~the operators either carry $\Delta B = \Delta L=0$ or  $\Delta B =\Delta L=1$; we focus on the former, the latter being discussed in Refs.~\cite{Heeck:2026dmh,Heeck:2026lhd}:
\begin{align}
\begin{split}
    \mathbb{O}_8 =
    \{
    & \bar{H} \bar{Q} \bar{L} e u H , \bar{H} \bar{Q} \bar{L} L Q H , \bar{H} \bar{Q} \bar{e} e Q H , \bar{d}e\bar{H}H\bar{L}Q,\\
    & \bar{H} \bar{d} \bar{L} L d H , \bar{H} \bar{d} \bar{e} e d H , \bar{H} \bar{u} \bar{L} L u H , \bar{H} \bar{u} \bar{e} e u H,\\
    & \bar{H} \bar{L} \bar{L} L L H , \bar{H} \bar{L} \bar{e} e L H , \bar{H} \bar{e} \bar{e} e e H , \bar{Q} \bar{L} e d H H,\\
    & \bar{u} \bar{L} e Q H H , \bar{u} \bar{L} L d H H , \bar{L} \bar{L} e e H H , \bar{H} \bar{H} \bar{L} e H H H,\\
    & \mathcal{D} \bar{Q} \bar{L} e Q H , \mathcal{D} \bar{Q} \bar{L} L d H , \mathcal{D} \bar{Q} \bar{e} e d H , \mathcal{D} \bar{u} \bar{L} L Q H,\\
    & \mathcal{D} \bar{u} \bar{e} e Q H , \mathcal{D} \bar{u} \bar{e} L d H , \mathcal{D} \bar{L} \bar{L} e L H , \mathcal{D} \bar{L} \bar{e} e e H,\\
    & \mathcal{D} \bar{d} \bar{L} e d H , \mathcal{D} \bar{u} \bar{L} e u H , \mathcal{D} \bar{H} \bar{H} \bar{L} L H H, \\
    & \mathcal{D} \bar{H} \bar{H} \bar{e} e H H, \mathcal{D} \mathcal{D} \bar{Q} \bar{L} e u, \mathcal{D} \mathcal{D} \bar{Q} \bar{L} L Q, \mathcal{D} \mathcal{D} \bar{Q} \bar{e} e Q, \\
    & \mathcal{D} \mathcal{D} \bar{Q} \bar{e} L d, \mathcal{D} \mathcal{D} \bar{d} \bar{L} L d, \mathcal{D} \mathcal{D} \bar{d} \bar{e} e d, \mathcal{D} \mathcal{D} \bar{u} \bar{L} L u, \\
    & \mathcal{D} \mathcal{D} \bar{u} \bar{e} e u, \mathcal{D} \mathcal{D} \bar{L} \bar{L} L L, \mathcal{D} \mathcal{D} \bar{L} \bar{e} e L, \mathcal{D} \mathcal{D} \bar{e} \bar{e} e e, \\
    & \mathcal{D} \mathcal{D} \bar{H} \bar{L} e H H, \mathcal{D} \mathcal{D} \mathcal{D} \bar{H} \bar{L} L H , \mathcal{D} \mathcal{D} \mathcal{D} \bar{H} \bar{e} e H
    \} \,.
\end{split}
\end{align}
Here, we are already omitting operators involving field-strength tensors, seeing as they have no tree-level UV completions~\cite{Li:2023pfw}. Some of these are instead studied in Refs.~\cite{PhysRev.126.375,Bowman:1978kz,Davidson:2020ord,Bryman:2021ilc,Fortuna:2022sxt,Fortuna:2023paj,Petrov:2013vka,Bhattacharya:2018ryy,Cai:2018cog}.
An explicit operator basis with well-defined index contractions was given by Murphy~\cite{Murphy:2020rsh} and is collected in Tab.~\ref{tab:Murphy_basis} for the non-derivative operators for the reader's convenience since we will refer to them below. An alternative basis can be found in Ref.~\cite{Li:2020gnx}.

We derived the UV completions for $\mathbb{O}_8$ using the code from Refs.~\cite{Heeck:2026dmh,Heeck:2026lhd}. We will not list them here since the non-derivative UV completions alone would take up 50 pages. All UV completions for $d\leq 8$ LFV SMEFT operators can be found as data files attached to the arXiv version of this article~\cite{anc_files} and agree with Ref.~\cite{Li:2023pfw} as far as we can tell. For non-derivative operators, which are phenomenologically more interesting, our files even include the induced operators as linear combinations of Murphy's basis operators. These can be obtained by integrating out the heavy fields from the Lagrangian and using Fierz and other identities to rewrite the operator in Murphy's basis. Ref.~\cite{Heeck:2026dmh} achieves this by solving a linear system in operator space without ever specifying identities, although we cross-checked many cases by hand to ensure correct results.

The majority of the \emph{derivative} $\mathbb{O}_8 $ UV completions are uninteresting in the sense that the mediator particles also induce lower-dimensional LFV operators, which then generically dominate the phenomenology.
In particular, the three-derivative $d=8$ operators are all UV-completed by the $d=6$ UV completions in Tab.~\ref{tab:d_6_der}, and are therefore inherently suppressed.
Similarly, almost all two-derivative operators arise by taking the appropriate $d=6$ UV completion of Tab.~\ref{tab:d_6} and expanding the mediator propagator to second order in momentum; fine-tuned cancellations aside, the two-derivative $d=8$ operators are then always suppressed compared to $d=6$. There is but one exception to this, already mentioned in Sec.~\ref{sec:UV6}: the $d=6$ operator $e HH\bar{H}\bar{L}$ can seemingly be UV-completed with two fermions $F_{(\vec{1},\vec{1},0)}$ \& $F_{(\vec{1},\vec{2},1/2)}$ via Fig.~\ref{Topo_6_2}, see Tab.~\ref{tab:d_6_Higgs_asymmetry}. But even though the Feynman diagram is valid, the corresponding amplitude simply vanishes to lowest order in the heavy-mediator expansion because it is antisymmetric under exchange of the two $H$. However, expanding the mediator propagator to second order in momenta gives a non-vanishing amplitude, so this is actually a UV completion for the $d=8$ operator $\mathcal{D}\mathcal{D}e HH\bar{H}\bar{L}$, specifically $\epsilon_{ij}\epsilon_{kl}\bar{L}_i H^\dagger_j H_k \mathcal{D}^2 (H^l e)$, containing $W$, $Z$, and the Higgs after electroweak symmetry breaking. This is the sole two-derivative $d=8$ UV completion that does not generate a $d=6$ operator at tree level. However, the required heavy singlet fermion $F_{(\vec{1},\vec{1},0)}$ by itself generates the $d=5$ Weinberg operator, which would then suppress the $d=8$ effects into oblivion, unless some form of total lepton number is imposed, say because neutrinos are Dirac particles.
 Aside from this one $\mathcal{D}^2$ example, this leaves only the one-derivative operators as potentially interesting. Using the code from Ref.~\cite{Heeck:2026lhd}, we constructed these one-derivative UV completions, which are attached to the arXiv version of this article~\cite{anc_files}.

\begin{table}[tb]
    \centering
    \renewcommand{\arraystretch}{1.5} 
    \setlength{\tabcolsep}{12pt} 
    \begin{tabular}{|c|c c|c|}
        \hline
        Operator & \multicolumn{2}{c|}{UV Completion}   & Topo. \\
        \hline \hline
        $eHH\bar{H}\bar{L}$
        & $F_{(\mathbf{1}, \mathbf{1}, 0)}$ & $F_{(\mathbf{1}, \mathbf{2}, \tfrac{1}{2})}$ & Fig.~\ref{Topo_6_2}\\
        \hline
    \end{tabular}
    \caption{Model that seemingly UV-completes the $d=6$ operator  $eHH\bar{H}\bar{L}$ but vanishes due to Higgs asymmetry and actually generates $\mathcal{D}\mathcal{D}eHH\bar{H}\bar{L}$ at $d=8$, see text for details.}
    \label{tab:d_6_Higgs_asymmetry}
\end{table}

\begin{figure}[tb]
    \centering
    \begin{subfigure}[b]{0.3\linewidth}
        \includegraphics[width=\linewidth]{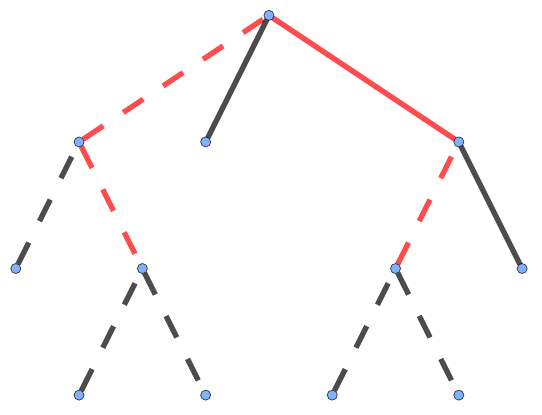}
        \caption{}
        \label{Topo_8_22}
    \end{subfigure}
    \hfill
    \begin{subfigure}[b]{0.3\linewidth}
        \includegraphics[width=\linewidth]{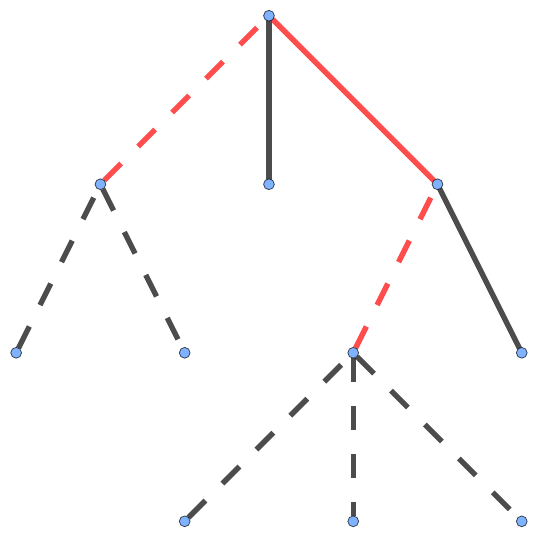}
        \caption{}
        \label{Topo_8_24}
    \end{subfigure}
    \hfill
    \begin{subfigure}[b]{0.3\linewidth}
        \includegraphics[width=\linewidth]{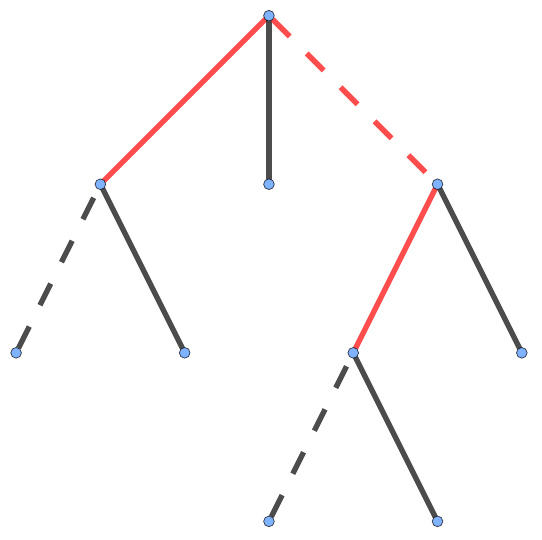}
        \caption{}
        \label{Topo_8_32}
    \end{subfigure}
    \caption{Topologies for tree-level UV completions of $d=8$ LFV operators, listing only those relevant for Tab.~\ref{tab:accidental_operators}.}
    \label{fig:topo32_group}
\end{figure}

Of course, $d=6$ LFV operators will \emph{always} be induced by the $d=8$ operators at loop level since they have the same quantum numbers, but at least at tree level one could hope to find UV completions that \emph{only} generate $d=8$ operators. Given our comprehensive lists of UV completions, it is straightforward to eliminate all $d=8$ mediators that would also induce $\mathbb{O}_{5,6,7}$, resulting in Tab.~\ref{tab:accidental_operators}. We refer to these as \emph{accidentally}-protected operators or scenarios, since it is the particle content that suppresses $d<8$ LFV rather than a symmetry as in Refs.~\cite{Heeck:2016xwg,Heeck:2024uiz,Heeck:2025jfs}.
The third column in Tab.~\ref{tab:accidental_operators} indicates the relevant topologies from Figs.~\ref{fig:dim7_topos} and~\ref{fig:topo32_group}. The fourth column gives the operator decomposed in Murphy’s basis~\cite{Murphy:2020rsh}, normalized such that the first operator in each decomposition carries a numerical prefactor of unity. The last column will be explained below.
As is obvious from the one-page table, only a tiny subset of $d=8$ operators and models can be accidentally protected, the vast majority generically induce LFV already at a lower mass dimension. This does not mean they are inherently uninteresting, but the subset from Tab.~\ref{tab:accidental_operators} comes with more motivation.
The accidentally protected operators fall into three distinct groups,  discussed in the next subsections. Notice that all four-fermion operators come from topology~\ref{Topo_8_32} and involve two quarks, which inevitably mix with the new colored fermions after electroweak symmetry breaking. These UV completions thus generically come with quark-flavor violation in addition to LFV, although we will not study these effects here.

\subsection{Non-derivative operators involving \texorpdfstring{$H$}{H} and \texorpdfstring{$H^\dagger$}{H-bar}}
\label{sec:H-Hbar}

The majority of operators in Tab.~\ref{tab:accidental_operators} look like $d=6$ operators multiplied by $H^\dagger H$. If  $H^\dagger H$ is indeed coupled to a total singlet, these operators become indistinguishable from $d=6$ operators at low energies, where $|H|^2\to v^2/2$. For example, the two UV completions for the operator $eHHH\bar{H}\bar{H}\bar{L}$ map onto Murphy's $\mathcal{O}_{LeH^5}$, which is just $|H|^2$ times the $d=6$ operator $\mathcal{O}_{e \phi}=(H^\dagger H)(\bar{L} H e)$~\cite{Grzadkowski:2010es}. Operators of this kind are marked by \check\, in Tab.~\ref{tab:accidental_operators} and are rather uninteresting. Note however that the $d=8$ UV completions will not only generate the $d=6$ operator structure at tree level, but also all other operators $\mathbb{O}_6$ at loop level, with potentially characteristic Wilson-coefficient ratios that could be used to disentangle these models from $d=6$ tree-level UV completions.

In other cases, $H^\dagger H$ is not coupled to a trivial singlet but rather to an $SU(2)_L$ triplet, so the relation to $d=6$ operators is slightly more complicated. For example, 
the UV completions of the operator $eH\bar{H}\bar{L}\bar{Q}u$ in Tab.~\ref{tab:accidental_operators} map onto linear combinations of the four operators $\mathcal{O}^{(j)}_{LeQuH^2}$, $j=1\ldots 4$, in Murphy's basis. The $j=1$ and $j=3$ operators are of the form $|H|^2 \mathcal{O}_{d=6}$, but $j=2$ and $j=4$ involve the Higgs coupling to a triplet, $H^\dagger \tau^I H$. As a result, the linear combinations contain the same terms as $d=6$ operators, but with different coefficients due to the different isospin structure. We mark such cases by (\check); in those cases, the phenomenology will still be similar to the $d=6$ case.

Let us mention one last special case in this category, the UV completion $F_{(\mathbf{3}, \mathbf{1}, -1/3)}$, $F_{(\mathbf{3}, \mathbf{3}, -1/3)}$, $S_{(\mathbf{3}, \mathbf{2}, -5/6)}$ for $H\bar{H}L\bar{L}Q\bar{Q}$. Since the scalar plus each fermion individually also leads to a realization of this operator with a different linear combination, listed in the row above and below this UV completion in Tab.~\ref{tab:accidental_operators}, the final operator depends on two complex coupling parameters. However, no coupling choices exist that would eliminate charged-LFV terms such as $\bar{e}\gamma^\mu P_L e\, \bar{q} \gamma_\mu P_L q$, so overall the phenomenology of this model will again be similar to the $d=6$ operator $L\bar{L}Q\bar{Q}$ up to isospin coefficients.
The mediator particles will furthermore generate other operators in Tab.~\ref{tab:accidental_operators}, leading to rich phenomenology.

\clearpage

\begin{widetext}

\begin{table}[H]
    \centering
    \newcolumntype{C}[1]{>{\centering\arraybackslash}m{#1}}
    \renewcommand{\arraystretch}{1.9} 
    \setlength{\tabcolsep}{5pt} 
    \begin{tabular}{|c|C{15mm}C{15mm}C{15mm}|c|c|c|}
        \hline
        Operator & \multicolumn{3}{|c|}{UV Completion}  & Fig. & Compact & $d=6$?\\
        \hline \hline
        $eHHH\bar{H}\bar{H}\bar{L}$
        &\multicolumn{3}{|c|}{
            $S_{(\mathbf{1}, \mathbf{3}, 0)} \quad \;
            S_{(\mathbf{1}, \mathbf{4}, \tfrac{1}{2})} \quad\;
            F_{(\mathbf{1}, \mathbf{4}, \tfrac{1}{2})} \quad\;
            S_{(\mathbf{1}, \mathbf{3}, 0)}$
        }&\ref{Topo_8_22} 
        & $\mathcal{O}_{LeH^5}$ & \check \\
        &$S_{(\mathbf{1}, \mathbf{4}, \tfrac{1}{2})}$&$F_{(\mathbf{1}, \mathbf{4}, \tfrac{1}{2})}$&$S_{(\mathbf{1}, \mathbf{3}, 0)}$
        &\ref{Topo_8_24}
        &$\mathcal{O}_{LeH^5}$ &  \check \\
        \hline
        $eH\bar{H}\bar{L}\bar{Q}u$
        &$F_{(\mathbf{3}, \mathbf{2}, \tfrac{1}{6})}$
        &$F_{(\mathbf{3}, \mathbf{3}, -\tfrac{1}{3})}$
        &$S_{(\mathbf{3}, \mathbf{3}, \tfrac{2}{3})}$
        &\ref{Topo_8_32}
        &$\mathcal{O}^{(1)}_{LeQuH^2} + \tfrac{1}{3} \mathcal{O}^{(2)}_{LeQuH^2} + \tfrac{1}{4} \mathcal{O}^{(3)}_{LeQuH^2} + \tfrac{1}{12} \mathcal{O}^{(4)}_{LeQuH^2}$ & (\check)\\
        &$F_{(\mathbf{3}, \mathbf{2}, \tfrac{7}{6})}$&$F_{(\mathbf{3}, \mathbf{3}, \tfrac{2}{3})}$&$S_{(\mathbf{3}, \mathbf{3}, \tfrac{5}{3})}$
        &\ref{Topo_8_32}
        &$\mathcal{O}^{(1)}_{LeQuH^2} - \tfrac{1}{3} \mathcal{O}^{(2)}_{LeQuH^2} + \tfrac{1}{4} \mathcal{O}^{(3)}_{LeQuH^2} - \tfrac{1}{12} \mathcal{O}^{(4)}_{LeQuH^2}$ & (\check)\\
        &$F_{(\mathbf{3}, \mathbf{1}, \tfrac{2}{3})}$&$F_{(\mathbf{3}, \mathbf{2}, \tfrac{7}{6})}$&$S_{(\mathbf{3}, \mathbf{1}, \tfrac{5}{3})}$
        &\ref{Topo_8_32}
        &$\mathcal{O}^{(1)}_{LeQuH^2} + \mathcal{O}^{(2)}_{LeQuH^2} + \tfrac{1}{4} \mathcal{O}^{(3)}_{LeQuH^2} + \tfrac{1}{4} \mathcal{O}^{(4)}_{LeQuH^2}$ & (\check)\\
        &$F_{(\mathbf{3}, \mathbf{2}, \tfrac{1}{6})}$&$F_{(\mathbf{3}, \mathbf{3}, -\tfrac{1}{3})}$&$S_{(\mathbf{3}, \mathbf{2}, -\tfrac{5}{6})}$
        &\ref{Topo_8_32}
        &$\mathcal{O}^{(1)}_{LeQuH^2} + \tfrac{1}{3} \mathcal{O}^{(2)}_{LeQuH^2} - \tfrac{1}{4} \mathcal{O}^{(3)}_{LeQuH^2} - \tfrac{1}{12} \mathcal{O}^{(4)}_{LeQuH^2}$ & (\check)\\
        &$F_{(\mathbf{3}, \mathbf{1}, -\tfrac{1}{3})}$&$F_{(\mathbf{3}, \mathbf{2}, \tfrac{1}{6})}$&$S_{(\mathbf{3}, \mathbf{2}, -\tfrac{5}{6})}$
        &\ref{Topo_8_32}
        &$\mathcal{O}^{(1)}_{LeQuH^2} - \mathcal{O}^{(2)}_{LeQuH^2} - \tfrac{1}{4} \mathcal{O}^{(3)}_{LeQuH^2} + \tfrac{1}{4} \mathcal{O}^{(4)}_{LeQuH^2}$ & \cross \\
        \hline
        $H\bar{H}L\bar{L}Q\bar{Q}$
        &$F_{(\mathbf{3}, \mathbf{3}, \tfrac{2}{3})}$&$F_{(\mathbf{3}, \mathbf{3}, \tfrac{2}{3})}$&$S_{(\mathbf{3}, \mathbf{4}, \tfrac{1}{6})}$
        &\ref{Topo_8_32}
        &$\mathcal{O}^{(1)}_{L^2Q^2H^2} + \tfrac{1}{3} \left(\mathcal{O}^{(2)}_{L^2Q^2H^2} + \mathcal{O}^{(3)}_{L^2Q^2H^2} + \mathcal{O}^{(4)}_{L^2Q^2H^2}\right)$ & (\check) \\
        &$F_{(\mathbf{3}, \mathbf{3}, -\tfrac{1}{3})}$&$F_{(\mathbf{3}, \mathbf{3}, -\tfrac{1}{3})}$&$S_{(\mathbf{3}, \mathbf{4}, -\tfrac{5}{6})}$
        &\ref{Topo_8_32}
        &$\mathcal{O}^{(1)}_{L^2Q^2H^2} - \tfrac{1}{3} \left(\mathcal{O}^{(2)}_{L^2Q^2H^2} - \mathcal{O}^{(3)}_{L^2Q^2H^2} + \mathcal{O}^{(4)}_{L^2Q^2H^2}\right)$ & (\check) \\
        &$F_{(\mathbf{3}, \mathbf{3}, -\tfrac{1}{3})}$&$F_{(\mathbf{3}, \mathbf{3}, -\tfrac{1}{3})}$&$S_{(\mathbf{3}, \mathbf{2}, -\tfrac{5}{6})}$
        &\ref{Topo_8_32}
        &$\mathcal{O}^{(1)}_{L^2Q^2H^2} + \tfrac{1}{3} \left(2 \, \mathcal{O}^{(2)}_{L^2Q^2H^2} - 2 \, \mathcal{O}^{(3)}_{L^2Q^2H^2} - \mathcal{O}^{(4)}_{L^2Q^2H^2}\right)$ & (\check) \\
        &$F_{(\mathbf{3}, \mathbf{1}, -\tfrac{1}{3})}$&$F_{(\mathbf{3}, \mathbf{3}, -\tfrac{1}{3})}$&$S_{(\mathbf{3}, \mathbf{2}, -\tfrac{5}{6})}$
        &\ref{Topo_8_32}
        &$\mathcal{O}^{(2)}_{L^2Q^2H^2} + \mathcal{O}^{(3)}_{L^2Q^2H^2} \pm \ii \,  \mathcal{O}^{(5)}_{L^2Q^2H^2}$ & (\check) \\
        &$F_{(\mathbf{3}, \mathbf{1}, -\tfrac{1}{3})}$&$F_{(\mathbf{3}, \mathbf{1}, -\tfrac{1}{3})}$&$S_{(\mathbf{3}, \mathbf{2}, -\tfrac{5}{6})}$
        &\ref{Topo_8_32}
        &$\mathcal{O}^{(1)}_{L^2Q^2H^2} + \mathcal{O}^{(4)}_{L^2Q^2H^2}$ & (\check) \\
        \hline
        $\bar{e}eH\bar{H}Q\bar{Q}$
        &$F_{(\mathbf{3}, \mathbf{3}, -\tfrac{1}{3})}$&$F_{(\mathbf{3}, \mathbf{3}, -\tfrac{1}{3})}$&$S_{(\mathbf{3}, \mathbf{3}, \tfrac{2}{3})}$
        &\ref{Topo_8_32}
        &$\mathcal{O}^{(1)}_{Q^2e^2H^2} - \tfrac{1}{3} \mathcal{O}^{(2)}_{Q^2e^2H^2}$ & (\check) \\
        &$F_{(\mathbf{3}, \mathbf{3}, \tfrac{2}{3})}$&$F_{(\mathbf{3}, \mathbf{3}, \tfrac{2}{3})}$&$S_{(\mathbf{3}, \mathbf{3}, \tfrac{5}{3})}$
        &\ref{Topo_8_32}
        &$\mathcal{O}^{(1)}_{Q^2e^2H^2} + \tfrac{1}{3} \mathcal{O}^{(2)}_{Q^2e^2H^2}$ & (\check) \\
        &$F_{(\mathbf{3}, \mathbf{1}, \tfrac{2}{3})}$&$F_{(\mathbf{3}, \mathbf{1}, \tfrac{2}{3})}$&$S_{(\mathbf{3}, \mathbf{1}, \tfrac{5}{3})}$
        &\ref{Topo_8_32}
        &$\mathcal{O}^{(1)}_{Q^2e^2H^2} - \mathcal{O}^{(2)}_{Q^2e^2H^2}$ & (\check) \\
        \hline
        $\bar{d}dH\bar{H}L\bar{L}$
        &$F_{(\mathbf{3}, \mathbf{2}, \tfrac{1}{6})}$&$F_{(\mathbf{3}, \mathbf{2}, \tfrac{1}{6})}$&$S_{(\mathbf{3}, \mathbf{3}, \tfrac{2}{3})}$
        &\ref{Topo_8_32}
        &$\mathcal{O}^{(1)}_{L^2d^2H^2} - \tfrac{1}{3} \mathcal{O}^{(2)}_{L^2d^2H^2}$ & (\check)\\
        \hline
        $\bar{d}d\bar{e}eH\bar{H}$
        &$F_{(\mathbf{3}, \mathbf{2}, \tfrac{1}{6})}$&$F_{(\mathbf{3}, \mathbf{2}, \tfrac{1}{6})}$&$S_{(\mathbf{3}, \mathbf{2}, -\tfrac{5}{6})}$
        &\ref{Topo_8_32}
        &$\mathcal{O}_{e^2d^2H^2}$ & \check \\
        &$F_{(\mathbf{3}, \mathbf{2}, -\tfrac{5}{6})}$&$F_{(\mathbf{3}, \mathbf{2}, -\tfrac{5}{6})}$&$S_{(\mathbf{3}, \mathbf{2}, -\tfrac{11}{6})}$
        &\ref{Topo_8_32}
        &$\mathcal{O}_{e^2d^2H^2}$ & \check \\
        \hline
        $H\bar{H}L\bar{L}\bar{u}u$
        &$F_{(\mathbf{3}, \mathbf{2}, \tfrac{1}{6})}$&$F_{(\mathbf{3}, \mathbf{2}, \tfrac{1}{6})}$&$S_{(\mathbf{3}, \mathbf{3}, \tfrac{2}{3})}$
        &\ref{Topo_8_32}
        &$\mathcal{O}^{(1)}_{L^2u^2H^2} + \tfrac{1}{3} \mathcal{O}^{(2)}_{L^2u^2H^2}$ & (\check)  \\
        &$F_{(\mathbf{3}, \mathbf{2}, \tfrac{7}{6})}$&$F_{(\mathbf{3}, \mathbf{2}, \tfrac{7}{6})}$&$S_{(\mathbf{3}, \mathbf{3}, \tfrac{5}{3})}$
        &\ref{Topo_8_32}
        &$\mathcal{O}^{(1)}_{L^2u^2H^2} - \tfrac{1}{3} \mathcal{O}^{(2)}_{L^2u^2H^2}$ & (\check)  \\
        &$F_{(\mathbf{3}, \mathbf{2}, \tfrac{7}{6})}$&$F_{(\mathbf{3}, \mathbf{2}, \tfrac{7}{6})}$&$S_{(\mathbf{3}, \mathbf{1}, \tfrac{5}{3})}$
        &\ref{Topo_8_32}
        &$\mathcal{O}^{(1)}_{L^2u^2H^2} + \mathcal{O}^{(2)}_{L^2u^2H^2}$ & (\check)  \\
        \hline
        $\bar{e}eH\bar{H}\bar{u}u$
        &$F_{(\mathbf{3}, \mathbf{2}, \tfrac{1}{6})}$&$F_{(\mathbf{3}, \mathbf{2}, \tfrac{1}{6})}$&$S_{(\mathbf{3}, \mathbf{2}, -\tfrac{5}{6})}$
        &\ref{Topo_8_32}
        &$\mathcal{O}_{e^2u^2H^2}$ & \check \\
        \hline
        \hline
        $deHH\bar{L}\bar{Q}$
        &$F_{(\mathbf{3}, \mathbf{2}, \tfrac{1}{6})}$&$F_{(\mathbf{3}, \mathbf{3}, -\tfrac{1}{3})}$&$S_{(\mathbf{3}, \mathbf{2}, -\tfrac{5}{6})}$
        &\ref{Topo_8_32}
        &$\mathcal{O}^{(3)}_{LeQdH^2} - \tfrac{1}{4} \mathcal{O}^{(4)}_{LeQdH^2}$  & \cross \\
        &$F_{(\mathbf{3}, \mathbf{2}, \tfrac{1}{6})}$&$F_{(\mathbf{3}, \mathbf{1}, -\tfrac{1}{3})}$&$S_{(\mathbf{3}, \mathbf{2}, -\tfrac{5}{6})}$
        &\ref{Topo_8_32}
        &$\mathcal{O}^{(3)}_{LeQdH^2} - \tfrac{1}{4} \mathcal{O}^{(4)}_{LeQdH^2}$ & \cross \\
        &$F_{(\mathbf{3}, \mathbf{2}, \tfrac{1}{6})}$&$F_{(\mathbf{3}, \mathbf{3}, -\tfrac{1}{3})}$&$S_{(\mathbf{3}, \mathbf{3}, \tfrac{2}{3})}$
        &\ref{Topo_8_32}
        &$\mathcal{O}^{(3)}_{LeQdH^2} + \tfrac{1}{4} \mathcal{O}^{(4)}_{LeQdH^2}$ & \cross \\
        \hline
        $dHHL\bar{L}\bar{u}$
        &$F_{(\mathbf{3}, \mathbf{2}, \tfrac{1}{6})}$&$F_{(\mathbf{3}, \mathbf{2}, \tfrac{1}{6})}$&$S_{(\mathbf{3}, \mathbf{3}, \tfrac{2}{3})}$
        &\ref{Topo_8_32}
        &$\mathcal{O}_{L^2udH^2}$ & \cross \\
        
                \hline\hline
        $\mathcal{D} \bar{H}H\bar{H}H\bar{L}L$
        & $F_{(\mathbf{1}, \mathbf{4}, \tfrac{1}{2})}$ & $S_{(\mathbf{1}, \mathbf{3}, 0)}$ & $S_{(\mathbf{1}, \mathbf{3}, 0)}$ & \ref{Topo_7_12}
        & $\mathcal{O}_{L^2H^4D}^{(4)}+ Y_e\mathcal{O}_{LeH^5}$ & \cross
\\    
        \hline
    \end{tabular}
    \caption{Accidentally protected $d = 8$ LFV operators. \check\, means the operator is just $|H|^2 \mathcal{O}_{d=6}$, making it phenomenologically uninteresting; (\check) means the operator is similar to one at $d=6$, but with different isospin factors. \cross\, means the operator is different from $d=6$ at tree level, e.g.~due to different chirality structure or missing neutral-current terms.
        }
    \label{tab:accidental_operators}
\end{table}
\end{widetext}

Finally, there is one case in this category with an interesting cancellation. Take the UV completion $F_{(\mathbf{3}, \mathbf{1}, -1/3)}$, $F_{(\mathbf{3}, \mathbf{2},1/6)}$, $S_{(\mathbf{3}, \mathbf{2}, -5/6)}$, i.e.~the last row of $eH\bar{H}\bar{L}\bar{Q}u$ in Tab.~\ref{tab:accidental_operators}. In unitary gauge, the induced operator becomes
\begin{align}
&\mathcal{O}^{(1)}_{LeQuH^2} - \mathcal{O}^{(2)}_{LeQuH^2} - \tfrac{1}{4} \mathcal{O}^{(3)}_{LeQuH^2} + \tfrac{1}{4} \mathcal{O}^{(4)}_{LeQuH^2} =\nonumber\\
&v^2 \left[ (\bar{\nu} P_R e) (\bar{d}P_R u) - \frac{1}{4}(\bar{\nu} \sigma_{\mu\nu}P_R e) (\bar{d} \sigma^{\mu\nu}P_R u)\right] \left(1+\frac{h}{v}\right)^2 \,,
\label{eq:cancellation}
\end{align}
in which all neutral-current pieces such as $(\bar{e}P_R e)(\bar{u}P_R u)$ cancel.\footnote{See Ref.~\cite{Gavela:2008ra} for a similar analysis of $d=8$ four-lepton operators.} This changes the phenomenology significantly, since the neutral-current terms typically give the strongest constraints via LFV signatures, and inevitably arise for $d=6$ SMEFT operators. Cases like this, which deviate from $d=6$ operators, are marked with \cross\, in Tab.~\ref{tab:accidental_operators} and provide the most interesting dimension-eight effects. The presence of neutrinos in the operator~\eqref{eq:cancellation} renders this operator less visible. By itself, it does not lead to charged LFV but rather modifications of charged-current interactions, e.g.~charged-current Non-Standard neutrino Interactions (NSIs) for first-generation quarks, specifically scalar and tensor NSI~\cite{Falkowski:2017pss,Falkowski:2019xoe,Falkowski:2019kfn,Coloma:2024ict}.
The dimension-eight origin of these NSI parameters leads to a scaling  $\epsilon \propto v^4/\Lambda^4\simeq 4\times 10^{-3} (\unit{TeV}/\Lambda)^4$,  $\Lambda$ being the SMEFT-scale of the $d=8$ operator. 
A large NSI coefficient~$\epsilon$ thus requires sub-TeV $\Lambda$. 
However, the same model also generates many other $d=8$ LFV operators, namely $d e H H \bar{L} \bar{Q}$ involving all mediators, $\bar{e} e H \bar{H} \bar{u} u$ and $\bar{d} d   \bar{e} e H\bar{H}$ through the scalar and $F_{(\mathbf{3}, \mathbf{2},1/6)}$, as well as $\bar{H} H L \bar{L} Q\bar{Q}$ through the scalar and $F_{(\mathbf{3}, \mathbf{1}, -1/3)}$. All of these generate charged LFV and, taken together, push $\Lambda$ above the TeV scale, so NSI will be heavily suppressed. $d=6$ LFV is generated at one-loop level by simply closing the Higgs loop to induce $e\bar{L}\bar{Q}u$, specifically the linear combination $\mathcal{O}^{(3)}_{lequ} - 4 \mathcal{O}^{(1)}_{lequ}$ in the notation of Ref.~\cite{Grzadkowski:2010es}, without any SM Yukawa suppression; if all mediator masses are degenerate $\equiv M$, the one-loop $d=6$ Wilson coefficient parametrically dominates over the tree-level $d=8$ one for $M > 4\pi v\simeq\unit[3]{TeV}$, but since the tree-level expression from Eq.~\eqref{eq:cancellation} involves a neutrino, the loop-level $e\bar{L}\bar{Q}u$ should give the better model constraints even for smaller $M$.
Overall, despite the curious cancellation in Eq.~\eqref{eq:cancellation}, this model is best probed through charged LFV, not NSI.

\begin{figure}[tb]
    \centering
 \includegraphics[width=0.9\linewidth]{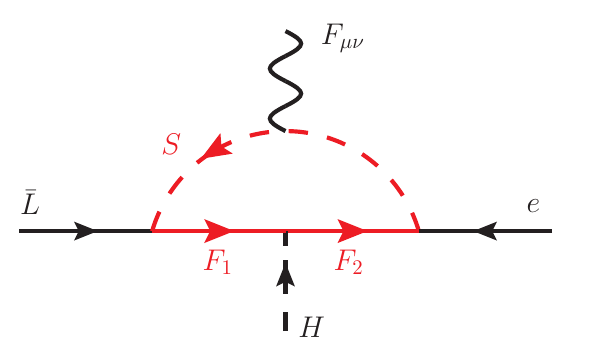}
    \caption{One-loop realization of the $d=6$ dipole operator $\bar{L}eH F$ with hypercharge or $SU(2)_L$ field strength $F_{\mu\nu}$ attached to the scalar line as an example.}
    \label{fig:d6_dipole}
\end{figure}

The particle content in the above example also induces one-loop $d=6$ dipole operators that are not suppressed by SM Yukawas, but are proportional to the coupling coefficient of $\bar{F}_{(\mathbf{3}, \mathbf{2}, 1/6)} H F_{(\mathbf{3}, \mathbf{1}, -1/3)}$ that does not contribute to the $d=8$ operators, as illustrated in Fig.~\ref{fig:d6_dipole}. Since there is no symmetry that could forbid this coupling, the model is expected to induce $\ell\to \ell' \gamma$ decays in addition to the processes induced by the semi-leptonic operators discussed in the previous paragraph, albeit proportional to independent overall coefficients.

\subsection{Non-derivative operators involving \texorpdfstring{$H^2$}{HH}}

Two of the operators in Tab.~\ref{tab:accidental_operators} contain four fermions and two $H$. Since $H^2$ is not a total gauge-group singlet, these operators are not valid $d=6$ operators if we stripped away the $H$. As a result, these $d=8$ operators, all marked with an \cross\, in Tab.~\ref{tab:accidental_operators}, are different from $d=6$ SMEFT operators~\cite{Ardu:2021koz}. At low energies, they still map onto four-fermion LEFT operators, just with different chirality structures compared to $d=6$ SMEFT operators. Indeed, $d=8$ SMEFT operators of this type are necessary to generate the full set of LEFT operators at tree level~\cite{Ardu:2021koz,Hamoudou:2022tdn}.

The operator $deHH\bar{L}\bar{Q}$ in Tab.~\ref{tab:accidental_operators} is induced in the linear combinations
\begin{align}
&\mathcal{O}^{(3)}_{LeQdH^2} \pm \tfrac{1}{4} \mathcal{O}^{(4)}_{LeQdH^2} \label{eq:deHHLQ}\\
&\  =\frac{v^2}{2} \left[ (\bar{e} P_R e)(\bar{d}P_R d) \pm \tfrac14 (\bar{e} \sigma_{\mu\nu} P_R e)(\bar{d}\sigma^{\mu\nu}P_R d)\right] \left( 1+ \frac{h}{v}\right)^2\nonumber
\end{align}
mapping onto the scalar and tensor LEFT operators $C_{S,LL}^{e\mu dd}$, $C_{S,RR}^{e\mu dd}$, $C_{T,LL}^{e\mu dd}$, $C_{T,RR}^{e\mu dd}$ in the notation of Ref.~\cite{Ardu:2021koz}.\footnote{A typo in Ref.~\cite{Ardu:2021koz} lists $ \mathcal{O}^{(4)}_{LeQdH^2}$ as $ \mathcal{O}^{(5)}_{LeQdH^2}$ in Eqs.~(67) \& (68) or Eqs.~(B.35) \& (B.36) in the published version.}
If one lepton is a muon and the other an electron, these operators or models are best searched for in $\mu$-to-$e$ conversion on nuclei~\cite{Ardu:2021koz}; for tauons, these operators would induce charged-LFV decays involving mesons. Compared to the previous subsection~\ref{sec:H-Hbar}, the Higgs loop cannot be closed here, so no $d=6$ operator proportional to the same couplings that enter $deHH\bar{L}\bar{Q}$ is generated. One can still attach an SM quark Yukawa and find a two-loop realization of the $d=6$ dipole LFV operators, but these parametrically very suppressed for TeV-scale mediators compared to~\eqref{eq:deHHLQ}.
Taking into account other Lagrangian terms, we note that these mediator particles also induce other operators in Tab.~\ref{tab:accidental_operators}, albeit not directly comparable to~\eqref{eq:deHHLQ}. In all three cases, we also find one-loop contributions to $d=6$ dipole operators proportional to the coupling coefficient of $\bar{F}_{(\mathbf{3}, \mathbf{2}, 1/6)} H F_{(\mathbf{3}, \mathbf{1} \text{ or } \mathbf{3}, -1/3)}$ as illustrated in Fig.~\ref{fig:d6_dipole}. 
Overall, the accidentally protected models behind $deHH\bar{L}\bar{Q}$ give rise to various LFV operators, with the $d=8$ ones~\eqref{eq:deHHLQ} potentially phenomenologically dominant depending on the coupling structure and mass hierarchy. 

The other operator in this category, $dHHL\bar{L}\bar{u}$, is induced by one Dirac fermion $F_{\left(\mathbf{3}, \mathbf{2}, 1/6\right)}$ and one scalar $S_{\left(\mathbf{3}, \mathbf{3}, 2/3\right)}$. It involves neutrinos and hence has not been discussed in the charged-LFV analysis of Ref.~\cite{Ardu:2021koz}. In unitary gauge, it reads
\begin{align}
\mathcal{O}_{L^2udH^2} = -\frac{v^2}{4} (\bar{e} \gamma^\mu P_L \nu) (\bar{u} \gamma_\mu P_R d)\left( 1+ \frac{h}{v}\right)^2 + \hc
\end{align}
and thus immediately maps onto the familiar vector-current charged-NSI coefficients $\epsilon^{d,u,P_R} \propto v^4/(8\Lambda^4)$, with $d=6$ LFV operators suppressed by loops and Yukawa couplings. Potentially testable NSI coefficients, $|\epsilon| \gtrsim 10^{-3}$~\cite{Ohlsson:2012kf,Farzan:2017xzy}, require $\Lambda \lesssim \unit[820]{GeV}$. However, the same mediator particles and couplings also generate $\bar{d} d H \bar{H} L \bar{L}$ and $H \bar{H} L\bar{L} \bar{u}u$, which together will push $\Lambda$ to much larger values through charged-LFV constraints. In addition, the relevant coupling $\bar{F}_{\left(\mathbf{3}, \mathbf{2}, 1/6\right)}S_{\left(\mathbf{3}, \mathbf{3}, 2/3\right)} L $ by itself generates the LFV $d=6$ dipole operators at one-loop level in the form
\begin{align}
    F_{\mu\nu}\,\bar{L}_p \sigma^{\mu\nu} \ii \slashed{\mathcal{D}} L_r\  \to\  (Y_e)_{rs}\bar{L}_p \sigma^{\mu\nu} H e_s\, F_{\mu\nu} \,,
    \label{eq:one-loop_dipole}
\end{align}
suppressed by an SM lepton Yukawa $Y_e$ from the equations of motion $\ii \slashed{\mathcal{D}} L_r = (Y_e)_{rs} H e_s $. Overall, the model behind $dHHL\bar{L}\bar{u}$ gives rise to many LFV channels that dominate over the original NSI operator.

\subsection{Derivative operators}

Finally, the one and only accidentally-protected derivative operator $\mathcal{D} \bar{H}H\bar{H}H\bar{L}L$ in Tab.~\ref{tab:accidental_operators} is proportional to
\begin{align}
    \overline{(H^\dagger \tau^I H) L_p} \mathcal{P}^{IJ}_{3/2}\ii \slashed{\mathcal{D}} (H^\dagger \tau^J H\, L_r)
    \label{eq:DHHHHLL}
\end{align}
with $SU(2)_L$ $\vec{3}\otimes \vec{2}\to \vec{4}$ projection operator $\mathcal{P}^{IJ}_{3/2} = \delta^{IJ} - \tfrac{1}{3}\tau^I \tau^J$ upon integrating out the heavy Dirac fermion $F_{(\mathbf{1}, \mathbf{4}, 1/2)}$ and (real) scalar $S_{(\mathbf{1}, \mathbf{3}, 0)}$; notice that one copy of the scalar is sufficient.
Using the product rule, integration by parts, and the equation of motion, Eq.~\eqref{eq:DHHHHLL} decomposes into the linear combination
\begin{align}
    \frac{1}{3}\mathcal{O}_{L^2H^4D}^{(4), pr} + \frac{1}{3} (Y_e)_{rs} \mathcal{O}_{LeH^5}^{ps}+ \frac{1}{3} (Y_e^\dagger)_{sp} (\mathcal{O}_{LeH^5}^{rs})^\dagger
\end{align}
with Murphy's basis operator
\begin{align}
\mathcal{O}_{L^2H^4D}^{(4),pr} &= \epsilon^{IJK} (\bar{L}_p \gamma^\mu \tau^I L_r)(H^\dagger \tau^J H) \mathcal{D}_\mu (H^\dagger \tau^K H)\\
&=\frac{g v^4}{2\sqrt2} \bar{\nu}_p \gamma^\mu P_L e_r\, W^+_\mu \left(1+\frac{h}{v}\right)^4 + \hc
\end{align}
Let us focus on $\mathcal{O}_{L^2H^4D}^{(4)}$, which involves one neutrino and thus does not generate charged LFV, but rather charged-current NSI upon integrating out the $W$-boson with modified coupling $\delta g^{W e}_L \propto v^4/(2\Lambda^4)$~\cite{Falkowski:2017pss,Falkowski:2019xoe,Falkowski:2019kfn,Coloma:2024ict} at low energies.
Notice that a similar operator arises at $d=6$, $O^{(3)}_{H\ell} = (\bar{L} \gamma^\mu \tau^I L)(H^\dagger \ii\, \raisebox{2mm}{\boldmath ${}^\leftrightarrow$}\hspace{-4mm} \mathcal{D}_\mu^{\,I} H) $, but in addition to the charged-current term it comes with $\bar{e} \gamma^\mu P_L e \, Z_\mu$ terms that lead to LFV in $Z$ decays. The $d=8$ operator is hence a much cleaner version of the charged-current NSIs.
In addition, the model induces the Yukawa-suppressed non-derivative operator $(Y_e)_{rs} \mathcal{O}_{LeH^5}^{ps}$ which, for diagonal $Y_e$ and $p\neq r$, gives rise to an off-diagonal term in the lepton mass matrix. Diagonalization subsequently induces an off-diagonal Higgs coupling~\cite{Harnik:2012pb} suppressed by the Yukawa coupling of the heavier lepton, say $Y_r$: $Y_{h\, pr} \propto Y_r v^4/\Lambda^4$. The smallness of the lepton Yukawas together with the difficulty in Higgs measurements makes the induced $h\to \bar{\ell}_p \ell_r$ decays less sensitive than the $W$-boson coupling shifts of $\mathcal{O}_{L^2H^4D}^{(4)}$.
Ultimately, both off-diagonal $W$ and $h$ couplings provide weaker constraints on the model than the one-loop contributions to $\ell_p\to \ell_r \gamma$ from the coupling $F_{\left(\mathbf{1}, \mathbf{4}, 1/2\right)}S_{\left(\mathbf{1}, \mathbf{3}, 0\right)} L $, analogous to Eq.~\eqref{eq:one-loop_dipole}, because these amplitudes do not depend on the $H^\dagger SH $ coupling that feeds into the $d=8$ operator~\eqref{eq:DHHHHLL} and is required to be small by electroweak precision data, seeing as it generates a vacuum expectation value for the triplet scalar $S$.
The model at hand therefore induces LFV at tree level through the $d=8$ operator $\mathcal{D} \bar{H}H\bar{H}H\bar{L}L$, but the phenomenologically dominant LFV effect arises at one loop through $d=6$ operators, notably the dipoles.

\section{Conclusions}
\label{sec:conclusion}

In this article, we studied lepton flavor violation from dimension-eight SMEFT operators through the lens of UV completions by heavy scalars and fermions. The impressive experimental sensitivity to charged LFV in principle allows detection of $d=8$ operators, the theoretical showstopper being the typically dominant and fairly similar $d=6$ operators. 
We have shown that $d=8$ operators could be interesting after all, since we can find UV completions that, at tree level, \emph{only} induce $d=8$ LFV operators, with $d=6$ LFV operators suppressed by loops. The tiny set of particularly interesting accidentally-protected dimension-eight LFV operators are marked with a \cross\, in Tab.~\ref{tab:accidental_operators}. At tree level, many of these operators do not give rise to charged LFV but rather charged-current NSI, although charged LFV is induced  by the same couplings in all cases and suppress NSI below testable rates.
As expected, \textit{loop-induced} $d=6$ LFV dipole operators often arise as the leading model prediction, but sometimes depend on independent coupling constants, allowing for tree-level $d=8$ operators to play the dominant role, for example in the form of the scalar and tensor LFV LEFT operators in Eq.~\eqref{eq:deHHLQ} not generated at tree level by dimension-six SMEFT.
We find that most accidentally-protected scenarios require colored mediator particles and generically lead to quark-flavor violation as well. A dedicated phenomenological study of these models including collider signatures will be performed elsewhere.

\section*{Acknowledgements}
We thank Diana Sokhashvili for discussions. 
We have made excessive use of Renato Fonseca's \texttt{Mathematica} packages \texttt{Sym2Int}~\cite{Fonseca:2017lem} and \texttt{GroupMath}~\cite{Fonseca:2020vke}.
This work was supported by the U.S.~Department of Energy under Grant No.~DE-SC0007974.
We acknowledge Research Computing at The University of Virginia for providing computational resources that have contributed to the results reported within this publication.
All tree-level UV completions for $d\leq 8$ LFV SMEFT operators can be found in the data files attached to the arXiv version of this article~\cite{anc_files}.


\bibliographystyle{utcaps_mod}
\bibliography{BIB.bib}

\end{document}